\documentclass[a4paper,11pt]{article}
\usepackage{pos}
\usepackage{overpic}
\usepackage{amssymb}
\usepackage{amsfonts}
\usepackage{url}

\usepackage{etoolbox}
\apptocmd{\thebibliography}{\setlength{\itemsep}{3pt}}{}{}
\title{Status of Ground-based and Galactic Gamma-ray Astronomy}
 \ShortTitle{Status of Ground-based and Galactic Gamma-ray Astronomy}

\author*[a,b]{A.M.W. Mitchell}
\affiliation[a]{Department of Physics, ETH Zurich\\
  CH-8093 Zurich, Switzerland}

\affiliation[b]{Friedrich-Alexander-Universit\"at Erlangen-N\"urnberg, Erlangen Centre for Astroparticle Physics, \\
Erwin-Rommel-Str. 1, D 91058 Erlangen, Germany}
\emailAdd{amitchell@phys.ethz.ch}
\emailAdd{alison.mw.mitchell@fau.de}

\abstract{

This conference proceedings is a write-up of the Gamma-ray Indirect rapporteur talk given at the 37$^{\rm{th}}$ International Cosmic Ray Conference (ICRC 2021). In contrast to previous ICRCs, this years edition was held in a fully virtual format, with dedicated discussion sessions organised around specific scientific themes. Many of these topics span the two categories of Gamma-ray Indirect (GAI) and Gamma-ray Direct (GAD), observations of gamma-rays by ground-based and space-based facilities respectively. To cover this organisation by topic in a coherent manner, this GAI rapporteur contribution focuses predominantly (but not exclusively) on Galactic gamma-ray astronomy, whereas the GAD rapporteur contribution focuses predominantly (but not exclusively) on Extra-galactic gamma-ray astronomy. 
In recent years, the field has seen enormous progress in both theory and observation, particularly in identifying PeVatrons (accelerators of Cosmic Rays to PeV energies), studies of particle escape from the accelerator, and detection of gamma-ray transients, especially gamma-ray bursts. 
}

\FullConference{37$^{\rm{th}}$ International Cosmic Ray Conference (ICRC 2021)\\
		July 12th -- 23rd, 2021\\
		Online -- Berlin, Germany}

\begin{document}
\maketitle

\section{Introduction}
\label{sec:intro}
%some general info on the ICRC format and contents here

The focus of ground-based gamma-ray astronomy lies in studying the origins of energetic Cosmic Rays (CRs) through the gamma-ray emission produced in the interactions of CRs with their environment. At energies above $\sim$1\,TeV, ground-based instruments are necessary due to the low rate of CRs, necessitating a large detection area.  
Since the first TeV sources were detected by ground-based facilities around 30 years ago the field has matured considerably, with over 200 Very-High-Energy (VHE, $\gtrsim$ TeV) sources currently known \cite{TeVCat}. 
Just within the last couple of years, considerable progress has been made, with the detection of gamma-rays above 1\,PeV for the first time; and from multiple sources. 

Studies of CRs have occupied generations of physicists for over a hundred years and continue to be a rich area of active research. The all particle Cosmic Ray spectrum follows approximately a power law with a spectral index of $\sim -2.7$ over many orders of magnitude in energy, yet there are two spectral features of key interest: the so-called knee, a spectral softening at $\sim3$\,PeV; and the so-called ankle, a spectral hardening at $\sim3$\,EeV \cite{PDG_CR}.
A straight-forward yet well-established conjecture is that the slight differences in slope; in the energy regions below the knee, between the knee and the ankle, and above the ankle; indicate that different source classes are responsible for the bulk of the CRs at different energies. Extra-galactic sources are held responsible for the CRs above the ankle, whilst a Galactic origin is assumed for CRs below the knee. The transition between Galactic and Extra-galactic sources is thought to occur between the knee and the ankle, yet conclusive evidence for Galactic accelerators reaching sufficient energies remains elusive \cite{HillasSNR}. 

Nevertheless, a coherent picture is emerging that will be explored in this proceedings, a write-up of the Gamma-ray Indirect rapporteur talk at the 37$^{\rm th}$ International Cosmic Ray Conference (ICRC); a conference series that has documented our progress in the understanding of Cosmic Rays ever since the first edition in 1947 \cite{ICRC}. 

\subsection{The First Virtual ICRC}

Due to the ongoing global pandemic, the 37$^{\rm th}$ ICRC, originally intended to be held in Berlin, Germany and organised by host institutes in Germany, took place in a fully virtual format. This led to several changes in the organisation with respect to previous years; most notably that contributions were not allocated scheduled times, but rather organised thematically into a series of discussion sessions. Talks ($\sim12$ min) and flash-talks ( $\sim3$ min for posters) were pre-recorded and uploaded in advance of the conference, such that the limited time spent in live sessions could be focused on questions and answers to individual talks and include more general discussions on a broader theme.  

The Gamma-ray Indirect (GAI) track received over 250 contributions, more or less evenly split between talks and posters, with GAI contributions featuring in 14 discussion sessions, 11 of which were joint topical discussions with contributions from other tracks. The Gamma-ray Direct (GAD) track received 110 contributions, also roughly evenly split between talks and posters. 

Traditionally, submissions on gamma-ray astronomy have been subdivided into the Gamma-ray Direct (space-based measurements) and Gamma-ray Indirect (ground-based measurements) tracks at the ICRC, which is a natural split at the technical level. 
Due to the organisation by scientific topic, however, strictly splitting each discussion session according to whether contributions were submitted to the GAD or GAI track tended to disrupt the conceptual flow of scientific argument. For this reason, the GAI rapporteur will cover the specifically GAI sessions and the Galactic joint sessions, whereas the GAD rapporteur will cover the specifically GAD sessions and the Extra-Galactic joint sessions. 

This proceeding is therefore organised as follows: 
section \ref{sec:groundexpts} covers the techniques employed by ground-based gamma-ray experiments and summarises those presented at the ICRC. Section \ref{sec:galactic} covers Galactic accelerators in depth, whilst section \ref{sec:escape} covers the escape of energetic particles from the accelerator and propagation in the surrounding medium. Gamma-ray bursts in the very-high-energy (VHE) regime as detected by ground-based facilities, is the only extra-galactic topic included and is covered in section \ref{sec:GRBs}. Finally, section \ref{sec:end} summarises with an outlook towards the anticipated future developments in ground-based gamma-ray astronomy.

\begin{figure}
    \centering
    \includegraphics[width=\textwidth]{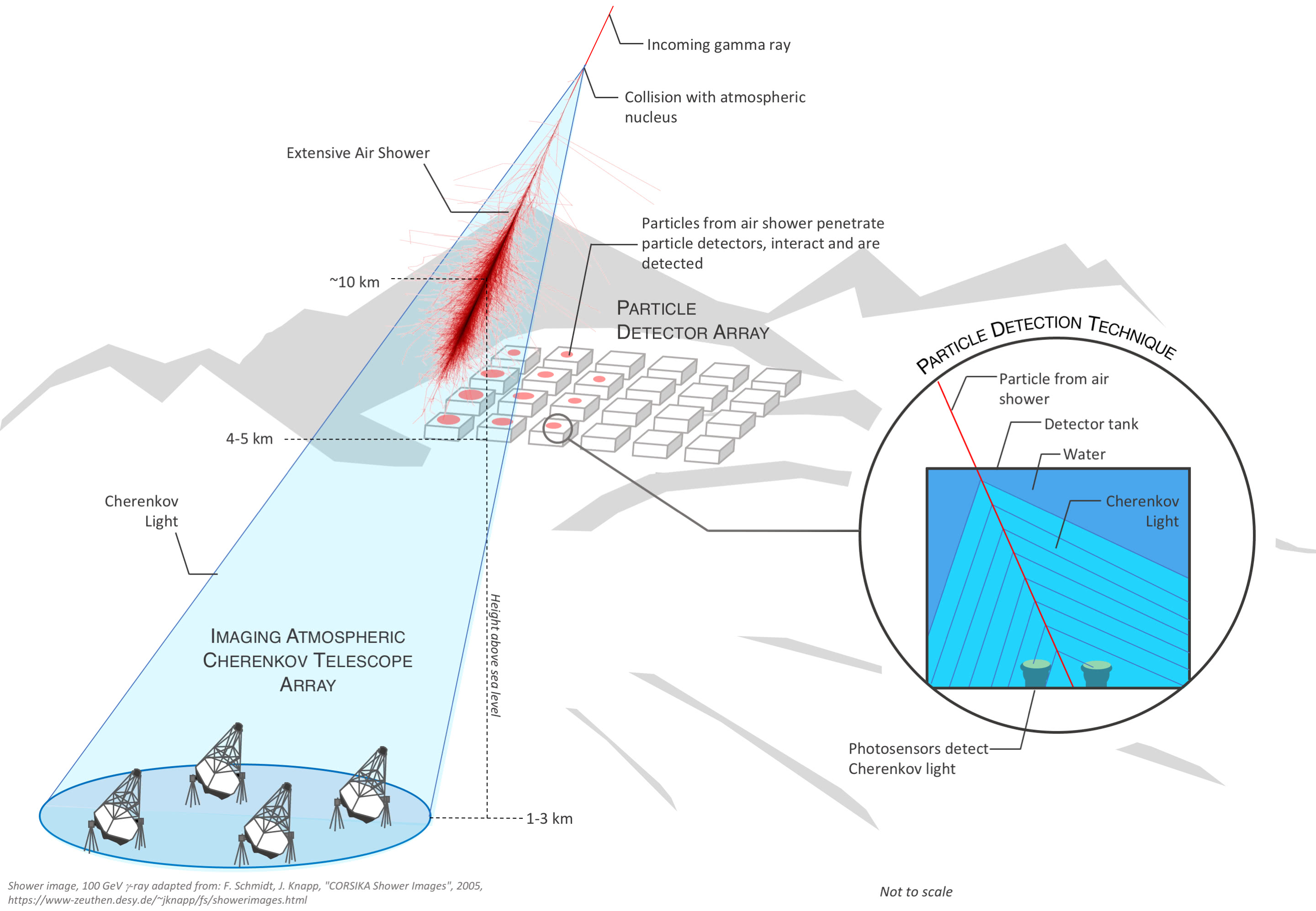}
    \caption{Schematic depiction of the two major detection methods for ground-based gamma-ray astronomy; particle detector arrays and Imaging Atmospheric Cherenkov Telescopes (IACTs) \cite{EASschematic}. Both techniques measure the Extensive Air Showers generated by incident gamma-rays.}
    \label{fig:EASschematic}
\end{figure}

\section{Ground-based Gamma-ray Facilities}
\label{sec:groundexpts}
%Overview of techniques and places

Incoming gamma-rays arriving at Earth interact with molecules in the atmosphere to initiate an Extensive Air Shower (EAS), a cascade of energetic particles and photons, through pair-production and Bremsstrahlung processes. %skip hadronic showers (?) OR mentino in detail. 
There are two main approaches to conducting gamma-ray astronomy from the ground, as shown in Figure \ref{fig:EASschematic}.  
Particle detector arrays situated at high altitudes of $\sim4-5$\,km\footnote{in order to be close to the altitude of shower maximum development typical in the TeV range} sample the particles in the EAS directly, for example through the Cherenkov light they generate when passing though water tanks at velocities exceeding the local speed of light, or from the luminescence signal generated by ionising radiation in a scintillator. 
Imaging Atmospheric Cherenkov Telescopes (IACTs) situated typically at altitudes of $\sim2000$\,m, just below the depth of shower maximum development, detect the Cherenkov light produced by the particles in the atmosphere \cite{EASschematic}.

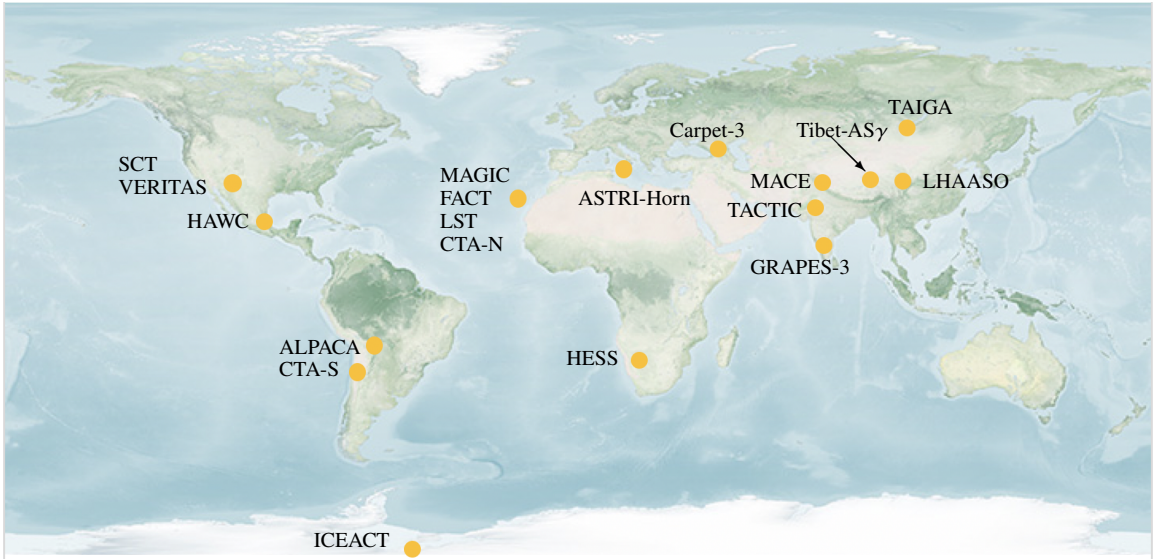
\begin{figure}
    \centering
%    \includegraphics{ICRC2021 template/figures/ICRC_GAI_WorldMap.png}
%grid
 \begin{overpic}[width=\textwidth]{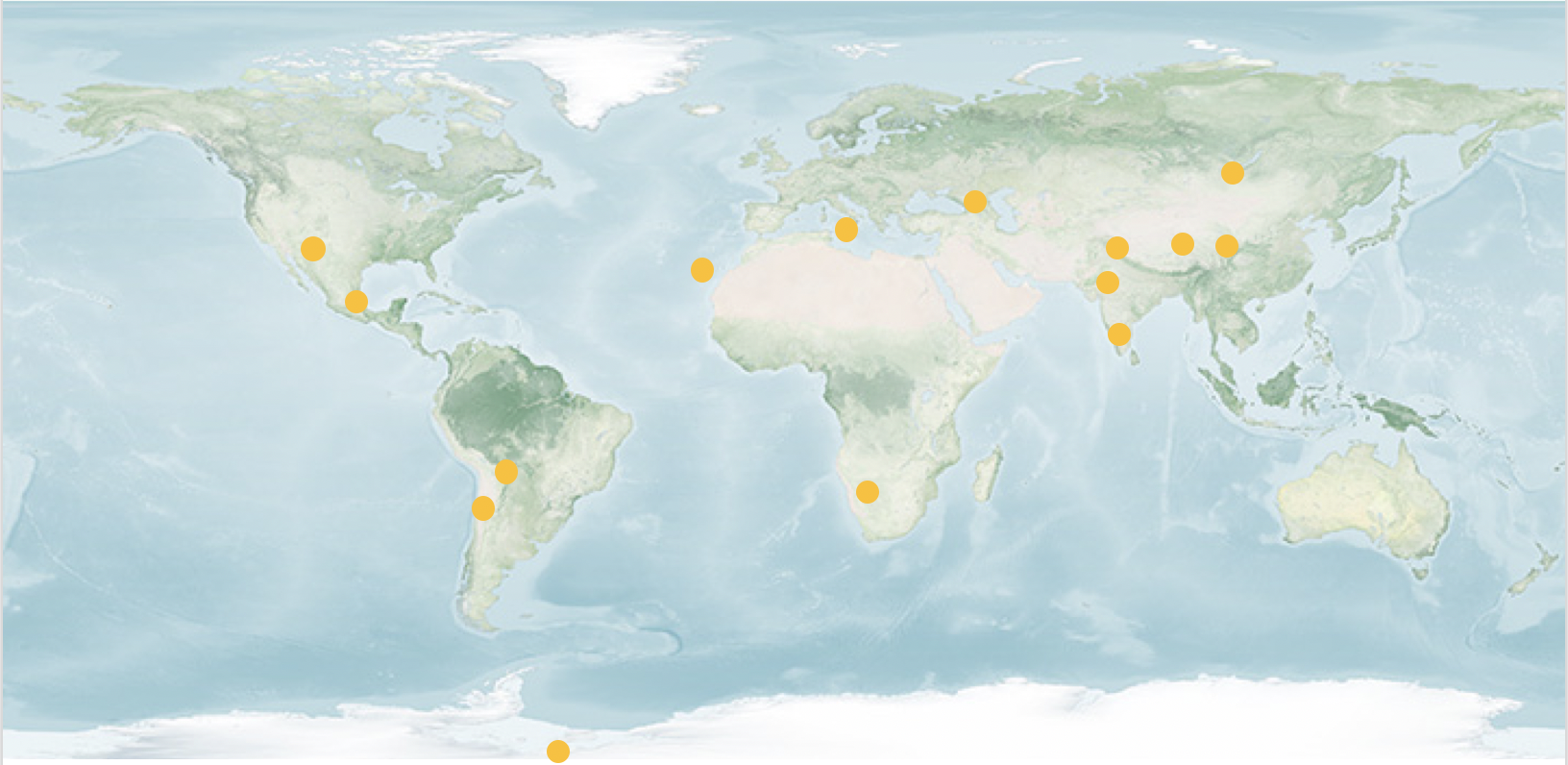}
    \put(10,32){\scriptsize{VERITAS}}
    \put(10,34){\scriptsize{SCT}}
    \put(16,29){\scriptsize{HAWC}}
    \put(24,18){\scriptsize{ALPACA}}
    \put(24,16){\scriptsize{CTA-S}}
    \put(27,1){\scriptsize{ICEACT}}
    \put(38,33){\scriptsize{MAGIC}}
    \put(38,31){\scriptsize{FACT}}
    \put(38,29){\scriptsize{LST}}
    \put(38,27){\scriptsize{CTA-N}}
    \put(50,31){\scriptsize{ASTRI-Horn}}
    \put(58,37){\scriptsize{Carpet-3}}
    \put(49,17){\scriptsize{HESS}}
    \put(63,30){\scriptsize{TACTIC}}
    \put(65,32.5){\scriptsize{MACE}}
    \put(65,25){\scriptsize{GRAPES-3}}
    \put(69,37){\scriptsize{Tibet-AS$\gamma$}}
    \put(72,37){\vector(1,-1){3}}
    \put(80,32.5){\scriptsize{LHAASO}}
    \put(77,39){\scriptsize{TAIGA}}
\end{overpic}
    \caption{World map indicating the locations of the current ground-based gamma-ray facilities featured at the ICRC. The properties of the different sites and experiments are summarised in table \ref{tab:GAI_expts} \cite{WorldMap}. }
    \label{fig:GAI_worldmap}
\end{figure}

\begin{table}
    \centering
    \begin{small}
    \begin{tabular}{l|cccc}
    Name & Location & Altitude a.s.l. (m) & Technology & Date \\
    \hline
    TAIGA \cite{TAIGA1,TAIGA2} & $51.81^\circ$N, $103.07^\circ$E & 675 & 2 IACTs $+$ AC array & 2018 (2014) \\
    VERITAS \cite{VERITAS} & $31.675^\circ$N, $110.95^\circ$W & 1268 & 4 IACTs & 2004 \\
    pSCT \cite{pSCT} & $31.675^\circ$N, $110.95^\circ$W & 1268 & 1 S-C IACT & 2019 \\
    TACTIC \cite{TACTIC} & $24.60^\circ$N, $72.70^\circ$E & 1300 & 1 IACT & 2001 \\
    Carpet-3 \cite{Carpet3} & $43.28^\circ$N, $42.69^\circ$E & 1700 & Scintillator array & 2021 \\
    ASTRI-Horn \cite{ASTRIhorn} & $37.70^\circ$N, $15.00^\circ$E& 1740 & 1 S-C IACT & 2019 \\
    HESS \cite{HESS} & $23.271^\circ$S, $16.5^\circ$E& 1800 & 4+1 IACTs & 2004 \\
    CTA-S \cite{CTA_RZ,CTA_OG} & $24.683^\circ$S, $70.316^\circ$W & 1800 & 14+40 IACTs & \textit{$\sim$2022} \\
    MAGIC \cite{MAGIC} & $28.762^\circ$N, $17.89^\circ$W & 2200 & 2 IACTs & 2004 \\
    FACT \cite{FACT} & $28.762^\circ$N, $17.891^\circ$W& 2200 & 1 IACT & 2011 \\
    CTA-N \cite{CTA_RZ,CTA_OG} & $28.762^\circ$N, $17.892^\circ$W & 2200 & 4+9 IACTs & \textit{$\sim$2022} \\
    LST \cite{LST} & $28.762^\circ$N, $17.892^\circ$W& 2200 & 1 IACT & 2018 \\
    GRAPES-3 \cite{GRAPES3} & $11.39^\circ$N, $76.66^\circ$E & 2200 & Scintillator array & 2000 \\
    IceACT \cite{IceACT} & $89.99^\circ$S, $63.453^\circ$W& 2840 & 2 ACT & 2019 \\
    HAWC \cite{HAWC} & $18.995^\circ$N, $97.309^\circ$W & 4100 & 300+345 WCDs & 2013 \\
    MACE \cite{MACE} & $32.78^\circ$N, $78.96^\circ$E & 4270 & 1 IACT & 2020 \\
    Tibet-AS$\gamma$ \cite{TibetAS} & $30.11^\circ$N, $90.53^\circ$E & 4300 & Scintillators+WCDs & 2014$^\dagger$ \\
    LHAASO \cite{LHAASO} & $29.359^\circ$N, $100.138^\circ$E & 4410 & Scintillators+WCDs & 2018\\ %2021 complete
    ALPACA \cite{ALPACA} & $16.383^\circ$S, $68.133^\circ$W & 4740 & Scintillator array & 2017$^*$\\
    \hline
    ASTRI \cite{ASTRI} & $28.30^\circ$N, $16.51^\circ$W & 2390 & 9 S-C IACTs & \textit{$\sim$2022}\\
    SWGO \cite{SWGO} & TBD & $>4500$ & WCDs & \\
    ALTO/CoMET \cite{ALTO,CoMET} & TBD & $>5000$ & Scintillators+CLiC & \\
    STACEX \cite{STACEX} & TBD & TBD & RPC & \\
    \hline
    \end{tabular}
    \end{small}
    \caption{Summary of the characteristics of current and future ground-based gamma-ray experiments. Their locations are indicated on the map in figure \ref{fig:GAI_worldmap}. The date provides the year (or expected year) in which the facility commenced regular data taking. Acronyms not previously defined include CLiC (Cherenkov Light Collectors), RPC (Resistive Plate Chambers) and AC (Non-imaging Air Cherenkov detectors).\\
    $^\dagger$ Tibet-AS has been operational since 1990, with the muon detector added in 2014. \\
    $^*$ ALPAQUITA, a small-scale part of ALPACA was constructed in 2017, with construction of half-ALPACA expected to be completed during 2022. }
    \label{tab:GAI_expts}
\end{table}

\begin{figure}
    \centering%lbrt
    \includegraphics[width=0.56\textwidth,trim=3cm 0mm 2cm 5mm,clip]{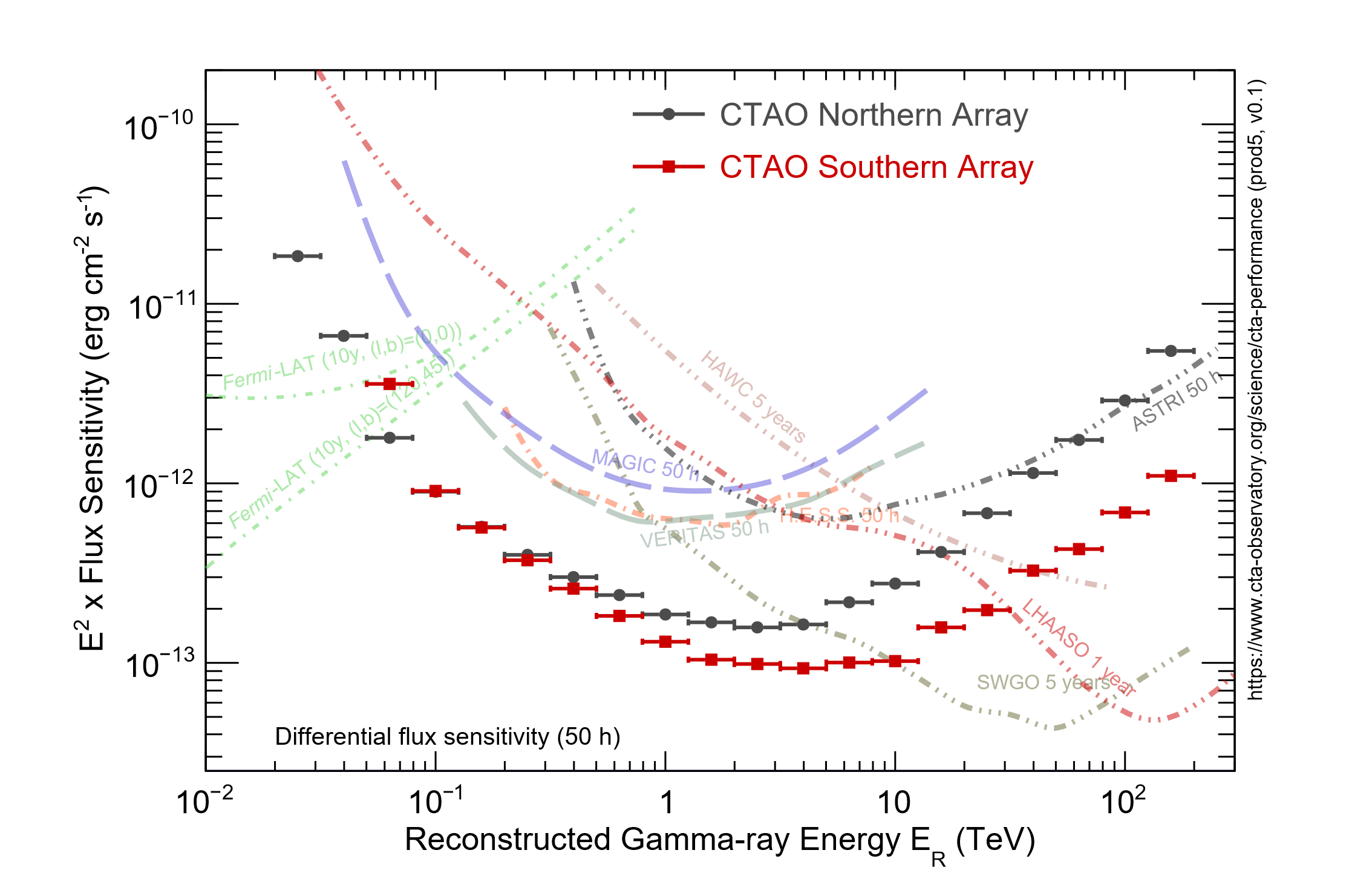}
    \includegraphics[width=0.43\textwidth,trim=3cm 0mm 2cm 5mm,clip]{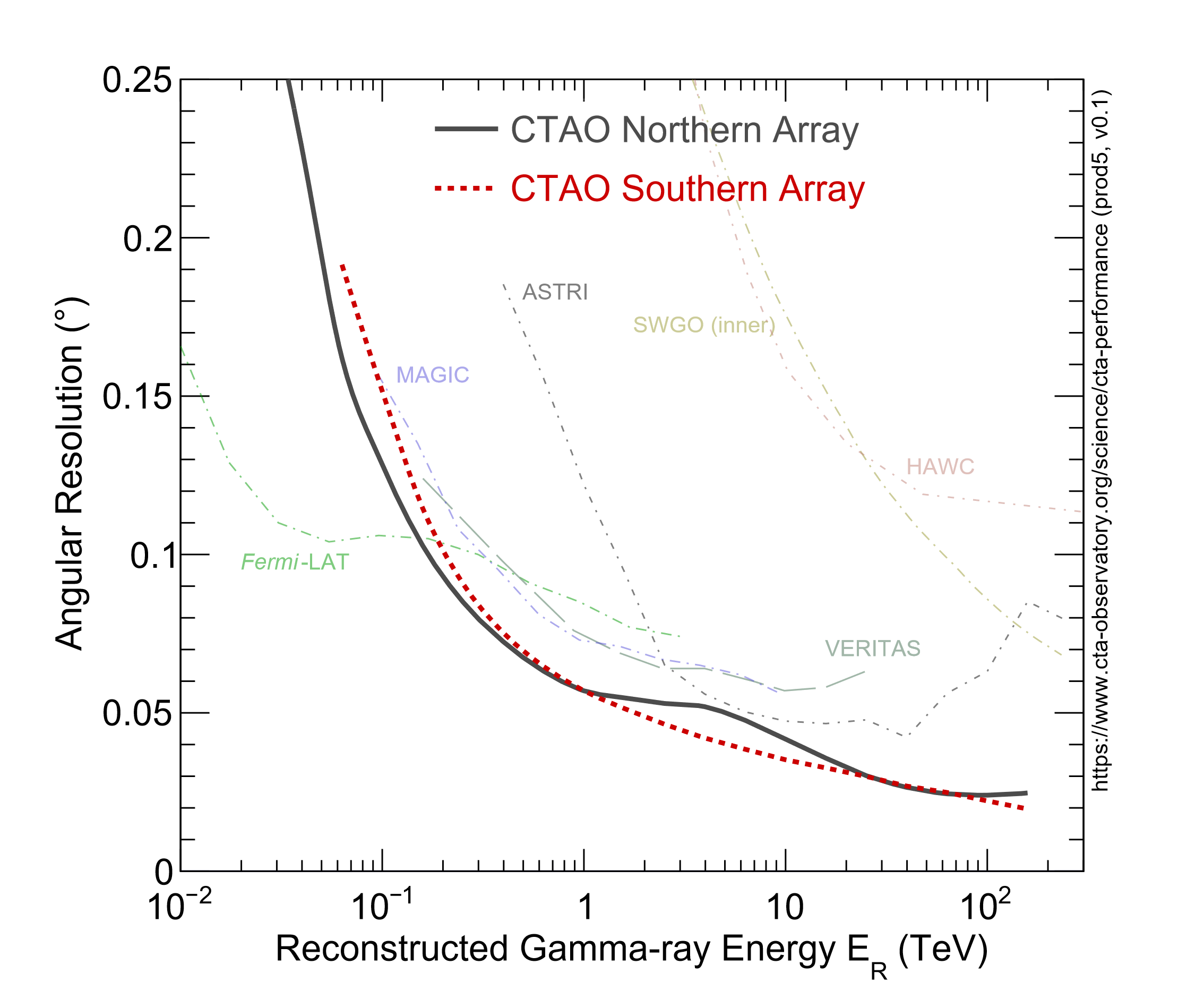}
    \caption{Performance curves for current (Fermi-LAT, HAWC, H.E.S.S., LHAASO, MAGIC, VERITAS) and future (ASTRI, CTA-N, CTA-S, SWGO) gamma-ray instruments. See Table \ref{tab:GAI_expts} for a summary of the different technologies used. Left: Sensitivity curves shown either for 50\,h (IACTs) or for a number of years observations (WCDs, particle detector arrays and the Fermi-LAT satellite). Right: angular resolution as a function of energy, typically defined as the opening angle of 68\% containment for the angular distance to the true direction \cite{CTAperf}. }
    \label{fig:CTA_perf}
\end{figure}

These two techniques are highly complementary, as illustrated in figure \ref{fig:CTA_perf}. In terms of flux sensitivity, IACTs perform better at energies $\lesssim1$\,TeV, whilst particle detector arrays have a much lower sensitivity at energies $\gtrsim10$\,TeV. The motivation for continuing to use both technologies up to 100s\,TeV is clearly demonstrated in the right hand panel of figure \ref{fig:CTA_perf}, which shows that the angular resolution of IACTs is substantially lower than that of particle detectors at all energies. 
IACTs therefore excel at precision studies, both in terms of source localisation and angular resolution, as well as spectral measurements and energy resolution. Nevertheless, IACTs have a limited duty cycle and can only operate at night time; preferably under low moonlight conditions in order to detect the faint Cherenkov light in the atmosphere. 

By contrast, particle detector arrays using closed water tanks or even underground detectors can in principle operate continuously and are able to observe a much wider region of the sky simultaneously, making them highly suited to monitoring and alerts. The field-of-view (FoV) is the angular region on the sky to which an instrument is sensitive at any given time; for IACTs, this is in the range of $\sim3^\circ - 10^\circ$, whilst for particle detector arrays this can be as much as 2\,steradian or 15\% of the sky \cite{HAWC}. 

Several facilities adopt a hybrid approach, using multiple technologies at a single site in order to combine their respective strengths to achieve the best performance. The transition between different constituent technologies providing the bulk of the performance at different energies accounts for some of the additional variation seen in figure \ref{fig:CTA_perf}, such as the `W' shape of the LHAASO sensitivity curve. In this case, the array of water Cherenkov detectors (WCDA) provides the sensitivity in the $0.1 - 10$\,TeV range, whilst the square kilometer array of scintillators (KM2A) provides the sensitivity $>10$\,TeV \cite{LHAASOsciasso}. 

%\red{mention fermi somewhere in context of hybrid measurements?}
Figure \ref{fig:CTA_perf} also includes the performance of Fermi-LAT, the only space-based gamma-ray detector to be included, which hence falls under the nominal category of gamma-ray direct (GAD). It is clear, however, that the sensitive energy ranges of ground-based instruments and Fermi-LAT overlap in the $\sim0.1-1$\,TeV range, such that gamma-ray measurements of astrophysical sources from both instruments are often combined to provide a more complete understanding. 

\subsection{Future Ground-based Gamma-ray Facilities}
\label{sec:future}
%future instrumentation

%The discussion session on \textit{New Instruments, Performance and Future Projects for Ground-Based Gamma-Ray Astronomy} provided an opportunity to highlight the work being done on instrument development. 

Many of the current experimental facilities are planning or have recently undergone technological upgrades, either to specific components (such as the cameras of IACTs, e.g. H.E.S.S. \cite{FlashCam}) or expanding the facility as a whole (such as the outriggers of HAWC \cite{outrigger}). 
In terms of future new-build facilities, it is clear that the geographical distribution shown in Figure \ref{fig:GAI_worldmap} is predominantly Northern hemisphere; which restricts access to the Southern sky. Several proposed future facilities (such as ALPACA, SWGO and ALTO/CoMET) included in table \ref{tab:GAI_expts} are therefore intended to be located in the Southern hemisphere, allowing the Southern sky - which includes the majority of the Milky Way and the Magellanic clouds - to be observed. 

\section{Galactic Gamma-ray Sources}
\label{sec:galactic}

As succinctly summarised in the introduction to the discussion session on \textit{The Origin of Galactic Cosmic Rays}, there are several clear stages that a particle undergoes to become a Cosmic Ray. 
The initial stage (i) is particle acceleration at the source, then (ii) the particles escape from their accelerator,followed by (iii) the propagation of the CRs across the Galaxy. Each one of these steps modifies the flux of CRs, with associated energy losses occurring for the second and third steps. 
This section will cover the first step, gamma-ray emission from Galactic sources as produced by accelerated CRs; the latter steps of particle escape and propagation are covered in section \ref{sec:escape}. 
Note here that these steps (i-iii) apply both to accelerators of hadronic CRs and leptonic CRs (electrons \& positrons); distinguishing their origins is a key open question. 

\subsection{The Galactic Source Population}
\label{sec:galpop}

The general view of the gamma-ray source population was addressed in the discussion session on \textit{The Census of Gamma-ray Sources}. 
Several contributions investigated the population of sources in the Galactic plane, looking at overall properties and the distribution between source classes. 
When comparing the number of sources (N) detected as a function of the source flux (S), commonly referred to as a logN-logS distribution, it is consistently found that too few point-like or low flux sources are recovered compared to model predictions \cite{Remy,Steppa}. This implies that there are likely many unresolved sources in the Galactic Plane, which may contribute to measurements of Galactic diffuse emission. %see later 

By comparing two different methods for source detection and identification, it could be shown in \cite{Remy} how two effects can lead to biased results when using a simple approach; namely a bias due to source confusion and a bias die to modelling. In the case of source confusion, the source flux is overestimated due to counting unresolved sources as part of a single, larger source, whereas in the case of modelling, the source flux is underestimated due to a single source with complex morphology being instead modelled as multiple smaller sources. 

Population synthesis approaches can be used to identify the contributions of different source classes to the total; however the source classes must be know apriori \cite{Steppa}. To identify new source classes in catalogues, first individual sources are studied in depth, with multi-wavelength information used as appropriate to clarify the source nature, before this information can be fed back into population studies. 
New source classes are thereby integrated into population models gradually, with each iteration leading to further refinement. %CUT some of this? 

%break
Despite numerous studies of Galactic gamma-ray sources, conclusively determining the origin of cosmic rays remains elusive. There are a number of eligible candidate source classes, including: Supernova Remnants (SNRs), Stellar Clusters, Superbubbles, Pulsars and Binary systems. 

\begin{figure}
    \centering
    \includegraphics[width=0.42\textwidth]{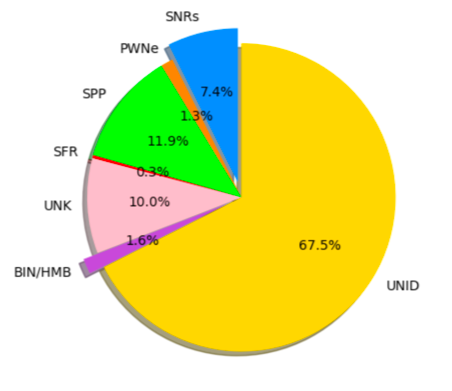}
    \includegraphics[width=0.42\textwidth]{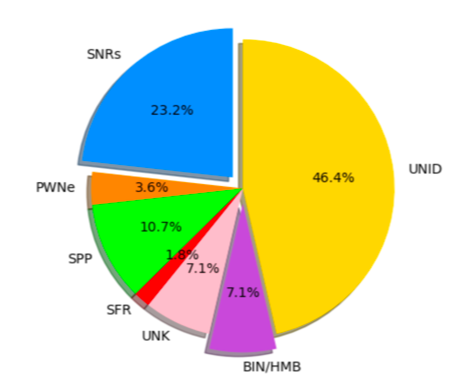}
    \caption{These pie charts show the proportion of Galactic gamma-ray sources belonging to a given source class in two samples. Left: 311 sources from the Fermi 4FGL catalogue used in the analysis. Right: 56 sources from the previous sample in which the pion bump feature was significantly detected \cite{LemoineGoumard}. }
    \label{fig:pionpie}
\end{figure}

%Move part to introduction? 
A key spectral signature of hadronic emission is the so-called pion-bump feature, that originates from the decay of energetic neutral pions into two gamma-rays. The `shoulders' of this feature occur at $\sim100$\,MeV-$1$\,GeV and $\sim1-10$\,TeV. 
Using sources listed in the 4FGL catalogue, \cite{LemoineGoumard} conducted a blind search for the lower energy shoulder of the pion bump. Figure \ref{fig:pionpie} shows their results: 311 sources passing their pre-selection criteria were analysed; 56 returned a significant detection of a pion-bump and are hence candidates for hadronic emission. Comparing the two distributions in figure \ref{fig:pionpie}, the proportion of SNRs and of binary systems notably increases in the pion-bump candidate sample, implying that these are potentially two hadronic source classes. 

Alternative sources of hadronic CRs, such as stellar clusters and superbubbles that have potentially many constituent parts are covered in section \ref{sec:windbubbles}. %cut part of sentence from 'that' to 'parts'?

\subsection{PeVatrons}
\label{sec:pevatron}

The dedicated discussion session on \textit{Ultra-High-Energy Gamma-Ray Sources and PeVatrons}, commenced with a deceptively simple question: ``What is a PeVatron?'' Usage in the literature has so far not converged to a unique definition, with the debate centering on two plausible alternatives: 
\begin{enumerate}
    \item \textit{An accelerator of hadronic cosmic rays to beyond 1\,PeV}
    \item \textit{An accelerator of particles (hadronic or leptonic) to beyond 1\,PeV.}
\end{enumerate}

The first definition is somewhat stricter as the gamma-ray emission must be proven to have a hadronic origin. 
For both definitions, ultra-high-energy (UHE) gamma-rays ($\gtrsim100$\,TeV) are a necessary, but not a sufficient condition for identification as a PeVatron. 
Presence of hadronic cosmic rays can be identified through an association with molecular clouds or by the rarer smoking gun signature of a coincident neutrino. 
It is worth bearing in mind, however, that a cloud illuminated by hadronic cosmic rays is evidence of PeVatron activity, but not in and of itself a PeVatron (accelerator). 
Therefore, a PeVatron ceases to be a candidate when a clear accelerator is identified with gamma-ray emission from energies beyond $\sim100$\,TeV. 
Which of the above two definitions is adopted is rather a moot point, but it is suggested that one is always clear in reference to `PeVatrons' in general, or `hadronic PeVatrons' / `leptonic PeVatrons' in particular. 

\begin{figure}
    \centering
    \includegraphics[width=\textwidth]{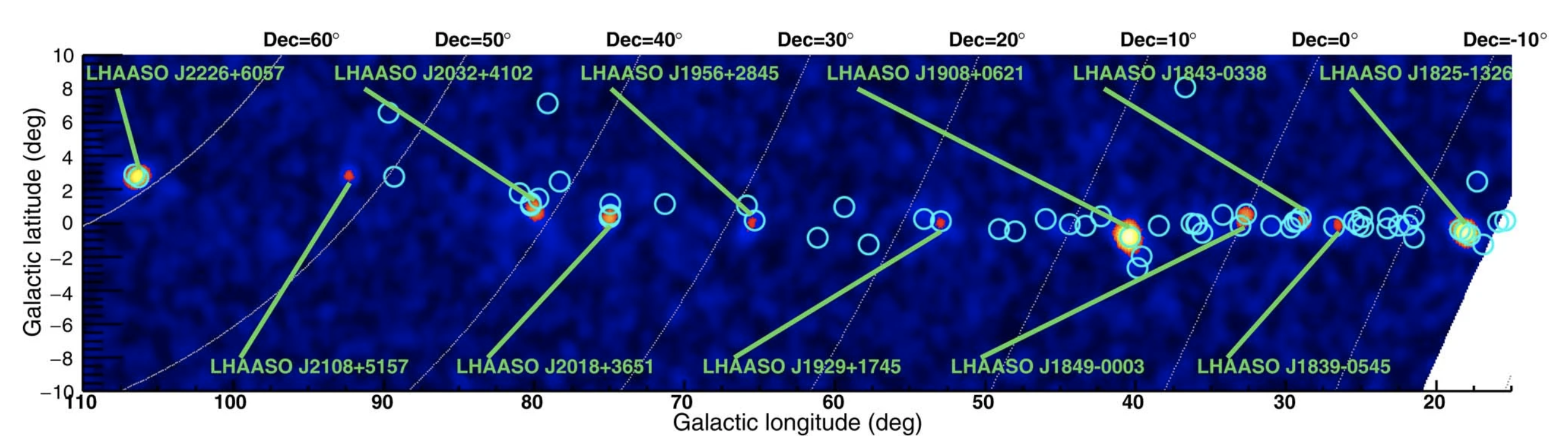}
    \caption{Map of the Galactic plane above 100\,TeV from LHAASO with 11 UHE sources shown \cite{LHAASOpevatrons}. Blue circles indicate the locations of known TeV sources.}
    \label{fig:lhaaso100}
\end{figure}

A spectacular result announced by the LHAASO collaboration earlier this year, was the identification of no fewer than 12 Galactic sources with UHE gamma-ray emission \cite{LHAASOpevatrons,LHAASOUHE}. This LHAASO view of the Galactic plane at UHE is shown in figure \ref{fig:lhaaso100}, whilst table \ref{tab:pevatron} lists the currently known UHE sources. In addition to those shown in figure \ref{fig:lhaaso100}, this list includes the Crab nebula \cite{LHAASOCrab}, the Galactic Centre region \cite{VeritasGC,HESSGC} and most recently HESS\,J1702-420 \cite{HESSJ1702}. 

\begin{table}
    \centering
    \begin{small}
    \begin{tabular}{c|ccc}
    Source & Location (l,b) & Detected $>100$\,TeV by & Possible Origin \\
    \hline 
    Crab Nebula & (184.557, -5.784) & {\footnotesize HAWC, MAGIC, LHAASO, Tibet-AS$\gamma$ }& PSR \\
    HESS\,J1702-420 & (344.304, -0.184) & {\footnotesize H.E.S.S.} & ? \\
    Galactic Centre & (0-1.2, -0.1-- +0.1) & {\footnotesize H.E.S.S.} & SMBH? \\
    eHWC J1825-134 & (18.116, -0.46) & {\footnotesize HAWC, LHAASO} & PSR \\
    LHAASO J1839-0545 & (26.49, -0.04) & {\footnotesize LHAASO} & PSR \\
    LHAASO J1843-0338 & (28.722, 0.21) & {\footnotesize LHAASO} & SNR \\
    LHAASO J1849-0003 & (32.655, 0.43) & {\footnotesize LHAASO} & PSR, YMC \\
    eHWC J1907+063 & (40.401, -0.70) & {\footnotesize HAWC, LHAASO} & SNR, PSR\\
    LHAASO J1929+1745 & (52.94, 0.04) & {\footnotesize LHAASO} & PSR, SNR\\
    LHAASO J1956+2845 & (65.58, 0.10) & {\footnotesize LHAASO} & PSR, SNR\\
    eHWC J2019+368 & (75.017, 0.283) & {\footnotesize HAWC, LHAASO} & PSR, H\,II/YMC\\
    LHAASO J2032+4102 & (79.89, 0.79) & {\footnotesize LHAASO} & YMC, PSR, SNR? \\
    LHAASO J2108+5157 & (92.28, 2.87) & {\footnotesize LHAASO} & ? \\
    TeV J2227+609 & (106.259, 2.73) & {\footnotesize Tibet-AS$\gamma$, LHAASO} & SNR, PSRs\\
    \end{tabular}
    \end{small}
    \caption{List of Galactic sources currently known to produce gamma-ray emission above 100\,TeV, together with their possible nature. Two sources have no known counterparts. }
    \label{tab:pevatron}
\end{table}

Several of the UHE sources listed in table \ref{tab:pevatron} exhibit energy-dependent morphology through the TeV range, or equivalently spectral variation across an extended source. Indications of this behaviour in J1908+063 were shown by both HAWC \cite{HAWCJ1908} and HESS \cite{HESSJ1908}, although neither can confirm this at a statistically significant level. HESS\,J1702-420 was demonstrated to shrink significantly at the highest energies; whilst the origin of the emission remains unknown, one possible explanation was speculated to be the existence of two separate components to the emission. It remains unclear if these components are two separate sources or intrinsically linked as part of the same gamma-ray source \cite{HESSJ1702}. 

TeV J2227+609 corresponds to the Boomerang nebula complex, with counterparts including SNR G106.3+2.7, (at an age $\geq$ 3.9\,kyr) and the highly energetic pulsar PSR\,J2229+6114 with a characteristic age $10$\,kyr and spin-down power $2.2\times10^{37}{\rm ergs}^{-1}$ \cite{ATNF}. %skip some? Explain pulsar properties here or explained later and cut here? 
CR interactions with molecular clouds in the region are plausible, yet the origin of the gamma-ray emission remains unclear. It is difficult to reconcile the SNR as the origin of the UHE emission given its developmental stage - the maximum energy of accelerated hadronic CRs should have already dropped below $\sim1$\,PeV, whilst the pulsar origin is energetically self-consistent \cite{OhnishiG106,OkaG106,LiuG106}.

On the theme of explaining the origin of UHE gamma-ray emission, it was previously thought that at energies beyond the Klein-Nishina limit, gamma-ray emission must be hadronic in origin, as the electrons undergo rapid, severe energy loss on each scattering interaction \cite{BlumenGould}. However, in \cite{Breuhaus} it is demonstrated that in high radiation environments, the maximum energy achieved through Inverse Compton scattering can be substantially increased to reach the UHE range. High radiation environments are defined through the balance of the radiation energy density $U_{\rm rad}$ and the magnetic energy density $U_B$, with $U_{\rm rad}/U_B = \Xi_{IC} \gg 1$ required to accelerate electrons beyond 100\,TeV \cite{Breuhaus}.

\subsection{Supernova Remnants}
\label{sec:snrs}

Supernova remnants (SNRs) are the canonical candidates for the origin of Galactic CRs and are believed to provide the bulk of the GCR flux. Nevertheless, substantial difficulties have been encountered (and substantial progress made) in the theoretical description of particle acceleration up to 1\,PeV in SNRs. 
Key themes of the dedicated session \textit{Supernova Remnants} included modelling of shell morphologies, especially descriptions of asymmetry, and of the SNR evolution; yet considerable emphasis was also placed in understanding the circumstellar environment, particle escape and particle transport (see section \ref{sec:escape}). 

Supernovae can be broadly categorised as one of two types; 
type II supernovae occur when a star reaches the end of its red giant phase having burnt through the available fuel. 
Type IA supernovae occur in binary systems, due to the accretion of material from a massive companion onto a white dwarf, until the Chandrasekhar limit is breached.\footnote{The Chandrasekhar mass limit of $\sim1.44M_\odot$ is the maximum stable mass that a white dwarf can support against gravitational collapse through electron degeneracy pressure.}
Therefore, type II events generally occur in younger systems, where the circumstellar and molecular environment is expected to be somewhat richer. 

Theoretical studies presented of the evolution of SNRs covered the remnant morphology, the maximum energy achieved by accelerated CRs, and the flux / luminosity evolution both in X-rays and gamma-rays. An exploration of the influence of the proper motion of the progenitor star on the resulting remnant morphology demonstrated pronounced asymmetry, dependent on the velocity, by using numerical 2D axisymmetric hydrodynamical simulations \cite{Meyer60M}. 
SN 1987A is the only supernova within living memory to have occurred within our Galaxy.\footnote{in the Large Magellanic Cloud, a satellite of the Milky Way at a distance of $\sim 50$\,kpc.} A dedicated model was developed \cite{Brose1987A} describing the flux evolution of the SNR and predicting that a VHE gamma-ray flux may become detectable within a few years time. 

The maximum energy (or momentum, $p_{\rm max}$) achieved by particles as a function of time was also explored theoretically for type II SNR with different progenitors \cite{MackeySNR} and for protons and electrons, where $p_{\rm max} = p_M(t/t_{\rm sed})^{-\beta}$ for times $t$ beyond the Sedov time $t_{\rm sed}$ at which the maximum momentum $p_M$ is achieved \cite{Celli}. The slope $\beta$ is a free parameter, typically $\sim2-4$.
For type II supernovae expanding into dense environments, a maximum achievable CR energy of $\sim100-200$ TeV within the first month post explosion was found \cite{MackeySNR}. For supernovae expanding into a uniform medium (type Ia) the maximum energy reached by protons is time-limited, whilst the maximum energy reached by electrons is loss-limited and affected by the CR-generated turbulent magnetic field \cite{Celli}.

Experimental results included two SNR with spectra that are characteristic of young SNRs: G150.3+4.5, which is measured in Fermi-LAT data in the GeV range as spectrally similar to other dynamically young and shell-type SNRs \cite{DevinG150}; and N132D situated in the LMC, for which the spectrum extends to beyond 10\,TeV, a feature expected rather for younger SNR than for N132D with its estimated age of $\sim2500$\,yr \cite{VinkN132D}. 

Detection of significant VHE gamma-ray emission from the Kepler's SNR was reported by H.E.S.S., completing the set of three historical SNRs with ages $\lesssim500$\,yr (along with Cassiopeia A and Tycho's SNR) as all shown to produce significant VHE emission \cite{ProkKepler}. 
Studies of two interacting SNRs were presented; W44 and G39.2-0.3 - in both cases, a hadronic scenario is a preferred fit to the spectral energy distribution, with evidence for dense molecular material nearby \cite{SushchG39}. In the W44 region, gamma-ray emission offset from the radio SNR shell and coincident with molecular material is thought to correspond to escaped CRs \cite{diVenereW44,PeronW44}. 

% \begin{figure}
%     \centering
%     \includegraphics[width=0.85\textwidth]{ICRC2021 template/figures/pwn_halo_sketch.pdf}
%     \caption{Caption \cite{GiacintiHalo} }
%     \label{fig:pwn_halo_sketch}
% \end{figure}

\begin{figure}
    \centering
    \includegraphics[width=0.5\textwidth]{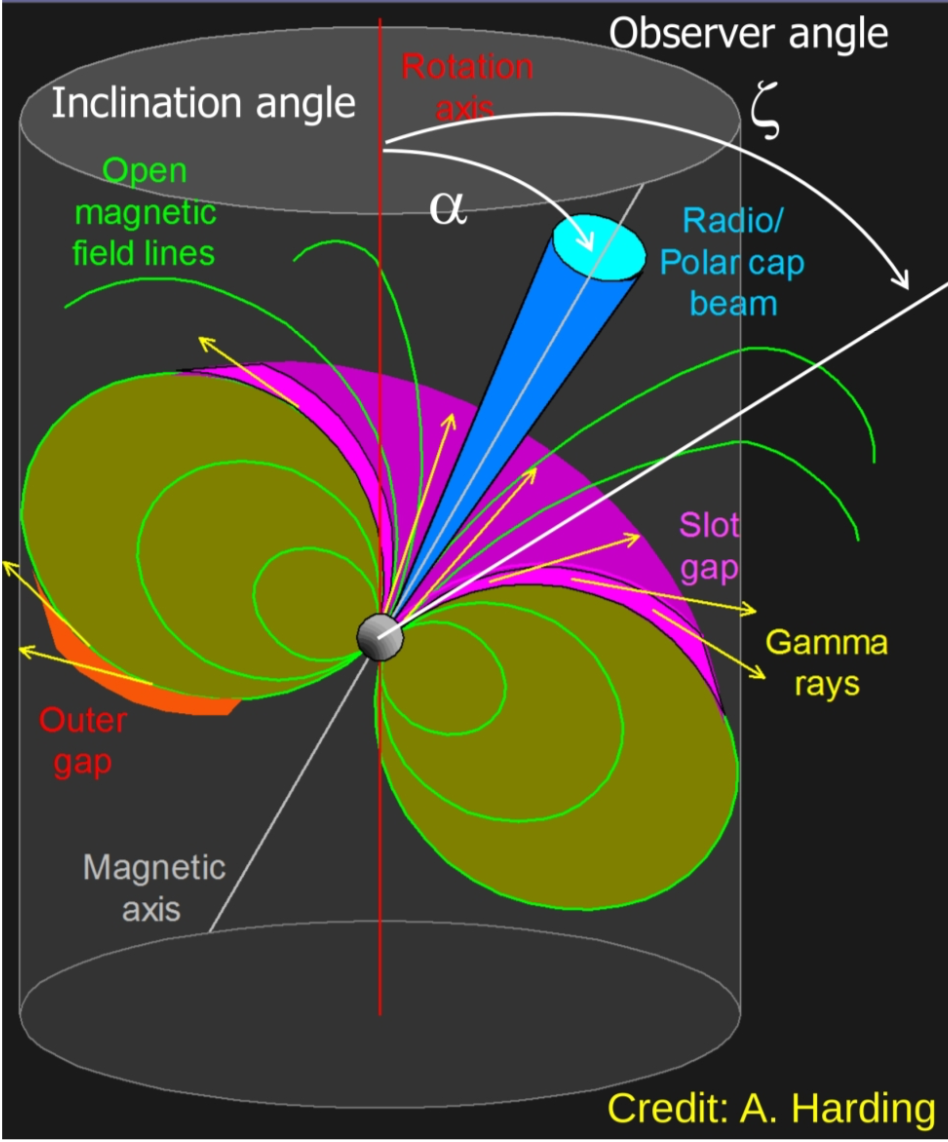}
    \caption{Schematic of the typical pulsar geometry - the magnetic field axis is inclined with respect to the rotation axis. Three plausible locations for efficient particle acceleration in the pulsar magnetosphere are considered in models: the polar cap, slot gap and outer gap. The cylindrical shaded area represents the light cylinder - at this radius, particles co-rotating with the pulsar would need to travel at the speed of light \cite{Barnard}. }
    \label{fig:pulsar_schematic}
\end{figure}

\subsection{Pulsars}
\label{sec:pulsars}

%Describe pulsar model of emission \red{swap order with PWN section?}
%As a supernova event occurs, the outer layers of the massive progenitor star are expelled outwards whilst the core will collapse in on itself and, depending on the mass, form either a White Dwarf, a Neutron Star or a Black Hole. During the collapse, angular momentum and magnetic flux are conserved, such that the degenerate remnant will be rapidly rotating with a strong magnetic field. Typically, the magnetic field axis need not be aligned with the rotation axis; due to the strong induced electric field, charged particles are taken from the surface of the neutron star and become trapped along the magnetic field lines. 

Pulsars featured as part of the discussion session on \textit{Galactic Compact Objects: Pulsars, Binary Systems, Microquasars}.
In the core collapse process accompanying a type II supernova event, a pulsar may form from the degenerate neutron star remnant, typically rapidly rotating with a strong magnetic field due to the conservation of angular momentum and magnetic flux during the collapse (see also figure \ref{fig:pulsar_schematic}). Charged particles extracted from the neutron star surface by the strong induced electric field become trapped along magnetic field lines. 

As the trapped particles co-rotate with the neutron star, synchrotron radiation is produced, with the photons subsequently producing electron-positron pairs. This pair-production process can lead to a pair multiplicity of $\sim 10^2 - 10^4$ pairs per primary particle \cite{Arons}. 
%leading to the formation of jets of emission along the magnetic field axis. If the cone of the beam sweeps past the line of sight to Earth, the emission will be seen as a pulsed signal; from which pulsars derive their name. 
At a distance from the surface of the star known as the light cylinder, charged particles must travel at the speed of light in order to continue co-rotating with the neutron star. The field lines become effectively open at this point and particles stream out in a magnetised pulsar wind. Exactly how this occurs remains uncertain, with three sites postulated for plausible particle escape, namely the polar cap, outer gap and slot gap (see figure \ref{fig:pulsar_schematic}).

Two contributions explored the influence of system geometry on the different components to the pulsar emission and attempted to fit VHE gamma-ray measurements of known pulsars \cite{Harding,Barnard}. To date, pulsed gamma-ray emission from four pulsars has been detected at VHE by IACTs. These are the Crab pulsar, Vela pulsar, Geminga pulsar and PSR\,B1706-44 \cite{MAGICCrab,HESSVelaPSR,MAGICGeminga,HESSB1706}.  

\begin{table}
    \centering
    \begin{tabular}{c|cccccc}
    Pulsar & PSR J-name& $P$ (ms) & $\tau_c$ (kyr)& $\dot{E}$ (erg s$^{-1}$) & $d$ (kpc) & VHE Ref. \\
         \hline
    Crab & PSR J0534+2200 & 33.4 & 1.26 & $4.5\times10^{38}$ & 2.0 & \cite{MAGICCrab} \\ %
    Vela & PSR J0835-4510 & 89.3 & 11.3 & $6.9\times10^{36}$ & 0.28 & \cite{HESSVelaPSR} \\% 
    Geminga & PSR J0633+1746 & 237 & 342 & $3.2\times10^{34}$ & 0.25 & \cite{MAGiCGeminga} \\ %
    PSR B1706-44 & PSR J1709-4429 & 102 & 17.5 & $3.4\times10^{36}$ & 2.6 & \cite{HESSB1706} %	
    \end{tabular}
    \caption{ The four pulsars for which VHE pulsed emission has been detected to date, listed in chronological order. Key properties taken from the ATNF catalogue \cite{ATNF} include the pulsar period $P$, the characteristic age $\tau_c$ (assuming magnetic dipole radiation and a braking index of $n=3$), the spin-down power $\dot{E}$ and the approximate distance $d$. Note that $\tau_c$ and $d$ are inherently uncertain parameters, whilst $P$ and $\dot{E}$ are well constrained. The pulsar J-name provides the location in RA-Dec coordinates. }
    \label{tab:pulsars}
\end{table}
%Geminga distance 0.19 in ATNF

Measurements of the VHE spectrum from the Geminga pulsar by MAGIC showed a power law extension of the emission up to 75\,GeV \cite{Ceribella}. Interpreting the emission in an outer gap scenario suggests that an Inverse Compton component is necessary to explain the highest energy pulsed emission. By contrast, using a polar cap model for the emission, \cite{Harding} were able to fully describe the emission using a single Synchro-Curvature component, with predictions for IC emission far below the sensitivity of ground-based facilities. 

Further interesting pulsar systems were presented using Fermi-LAT data; such as PSR\,J2021+4026 that exhibits apparent gamma-ray flux variability coincident with changes in the spin-down rate \cite{FioriPSR}. Such pulsar glitches, as they are commonly known, are thought to be linked to the star quakes on the pulsar surface and to a reorganisation of the magnetic field structure - an associated gamma-ray flux change suggests that the GeV emission is probing the pulsar magnetosphere; however more sophisticated models are desirable to further our understanding here.

Significant GeV emission has been detected from over 20 Globular Clusters to date, yet from only one at TeV energies, Terzan\,5 \cite{TeVCat}. As Globular Clusters are collections of older stars, collective emission from a population of millisecond pulsars is one of the more popular explanations, although the origin of the GeV emission has not yet been conclusively proven. 
Two contributions tested this scenario; by investigating correlations of the gamma-ray luminosity with astrophysical parameters of the clusters \cite{DSong}; and by comparing measured upper limits to the predicted gamma-ray flux under different model scenarios \cite{CVenter}. A mild tension is seen only in the case of the stacked upper limit, which the first combination of model parameters violated. Clearly further observations and detailed modelling are needed to constrain the origin of the gamma-ray emission from globular clusters.

\subsection{Pulsar Wind Nebulae}
\label{sec:pwne}

Particles streaming away from the light cylinder (see figure \ref{fig:pulsar_schematic}) form a relativistic pulsar wind. 
Where the pulsar wind is decelerated to match the slower expansion of ambient material within the nebula, a wind termination shock forms, %(at a radius where the ram pressure of the wind balances the internal pressure of the PWN)
at which further acceleration of the electrons/positrons occurs \cite{GaenslerSlane}. Within the pulsar wind, particles flow along magnetic field lines without radiating. Downstream of the wind termination shock, the re-accelerated particles produce synchrotron emission and form the pulsar wind nebula - a `bubble' of relativistic particles that forms initially around a pulsar and typically expands within an SNR \cite{GaenslerSlane}. 

As the forward shock of the PWN expands, it may encounter the reverse shock of the SNR, leading to shock mixing and a disrupted system. At this point the nebula `bubble' effectively `bursts' and particles are free to stream or diffuse out of the original nebula and disperse into the interstellar medium (ISM). These escaped particles may form a halo around the original PWN (see section \ref{sec:halos}), which has recently been classified as a distinct source class of energetic particles producing non-thermal emission through IC scattering in the ISM \cite{LindenHalo,GiacintiHalo}.

It is clear that the process of particles escaping the PWN is not instantaneous; there exists a transition period during which various proportions of the total particle population may be either contained within the PWN or escaped into the ISM respectively \cite{GiacintiHalo}. 
The manner in which sources should be classified as either PWN or halo remains a point of debate, with two observational methods proposed: 
\begin{enumerate}%[roman]
    \item diffusive particle transport preferred over advection as best-fit to the radial profile of the gamma-ray emission, when interpreted as particle escape. 
    \item particle energy density derived from the gamma-ray emission is $<0.1$eV/cm$^3$, far below the level of the ISM.
\end{enumerate}

It should be noted, however, that it is well-established that diffusive particle transport may dominate over advection also within the PWNe, such as the Crab nebula, although in most cases a combination of transport processes are likely at work \cite{Tang}.

\subsection{Binary Systems}
\label{sec:binaries}

%\red{Spider binaries / black widow \cite{SpiderBin}.}
`Spider' binaries are systems in which a millisecond pulsar is interacting with a low mass stellar companion. %Two subclasses of `black widows' and `redbacks' are distinguished by the mass of the stellar companion. 
The pulsar wind irradiates and effectively engulfs the stellar companion, which provides a rich target photon field for non-thermal emission. A modelling study \cite{SpiderBin} demonstrated that an enhanced and potentially detectable gamma-ray flux from these systems can be expected when in an optical flaring state. %These systems are therefore a promising new class of gamma-ray binaries for studies of oblique shock acceleration. 

Microquasars are binary systems comprised of a stellar mass black hole and a massive ordinary star. As matter is accreted from the companion star to the black hole, an accretion disk is formed together with jets along the rotation axis. Such microquasar systems are therefore like miniature versions of quasar Active Galactic Nuclei (AGN), whence their name. 

SS\,433 is an example of such a binary system, with gamma-ray emission detected by HAWC not from the microquasar itself, but from the jets that it produces \cite{SS433HAWC}. %This jet emission is significantly detected by HAWC; 
Using the HAWC emission as a template for the expected morphology, \cite{SS433Fang} recovered a significant ($>5\sigma$) signal in the Fermi-LAT GeV data corresponding to the jets. 
Alongside this signal coincident with the jets, a nearby GeV source Fermi\,J1913+0515 was detected. 
%preferred two source over single model

\begin{figure}
    \centering %lbrt
    \includegraphics[trim=2mm 2mm 0mm 0mm,clip,width=0.75\textwidth]{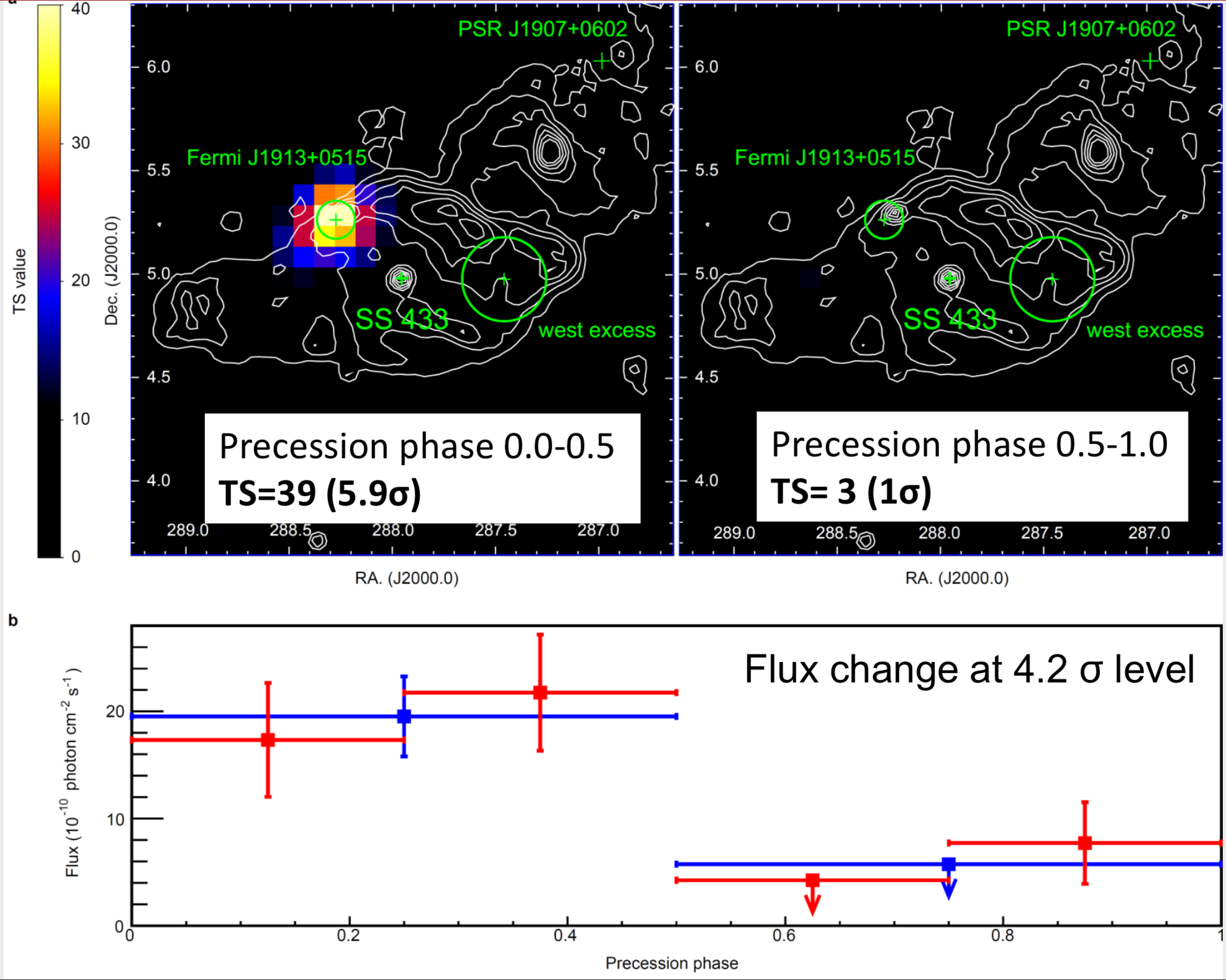}
    \caption{Fermi\,J1913+0515, a GeV bright region close to SS\,433, not in the line of the jets but coincident with an adjacent cloud, exhibiting a periodicity synchronous with the precession phase \cite{SS433JLi,SS433Nature}. }
    \label{fig:SS433}
\end{figure}

Fermi\,J1913+0515 is intriguing in its own right - somewhat remarkably, as shown in figure \ref{fig:SS433}, this source exhibits a periodicity to the gamma-ray flux with a period corresponding to the precession of the SS\,433 system. Such behaviour may be expected from the jets; yet the angle of the jets does not intersect the location of Fermi\,J1913+0515. Instead, the primary feature of note at the location of this periodic source is a molecular cloud \cite{SS433JLi}. To have a periodic gamma-ray signal from a cloud illuminated by CRs is highly surprising. 
%astonishing - it is an example of the surprises nature likes to throw our way; a challenge to our understanding. 
We can look forward to further developments in our knowledge of microquasar systems and to future insights with anticipation. 

\section{Particle Escape and Propagation}
\label{sec:escape}

Following the acceleration of CRs at their source, the CR spectrum is subsequently influenced by energy losses via the particle escape and propagation processes. %that will be addressed in this section. 
Indeed, it has been conjectured \cite{HillasGCR} that the `knee' of the CR spectrum does not necessarily signify a change of accelerator or source class, but could rather reflect an escape process or energetic limit for the confinement of particles to their accelerator. 
Escaping particles then propagate through the ISM, along trajectories modified by magnetic fields, and may interact with interstellar clouds. 

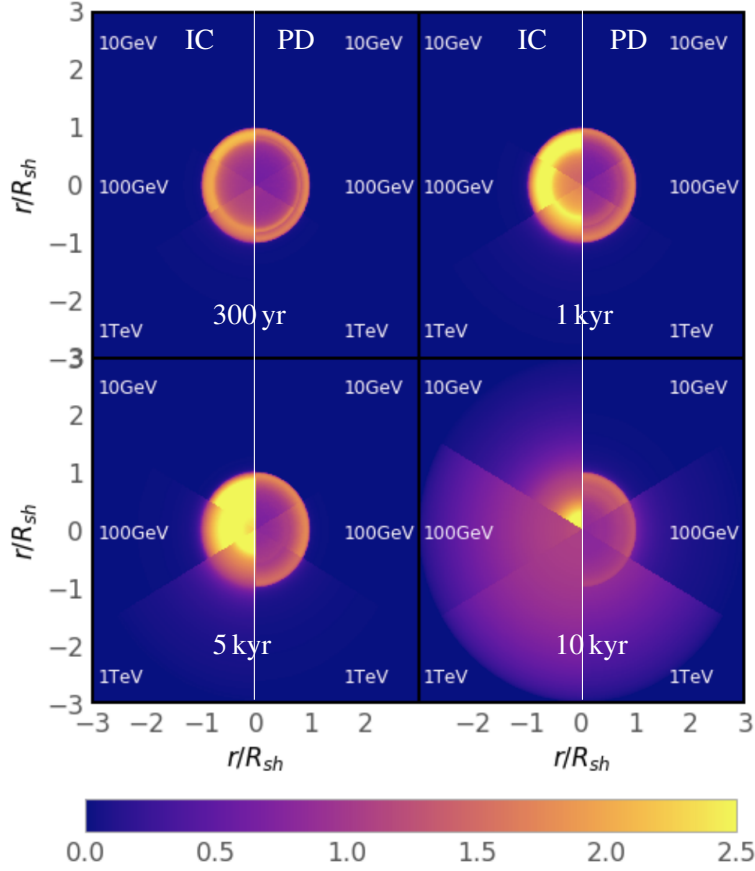
\begin{figure}
    \centering
    %Maybe only the second figure?!
    % \includegraphics[width=0.72\textwidth]{ICRC2021 template/figures/Brose_Diffusion_SNRhalo.png}
    \begin{overpic}[width=0.72\textwidth]{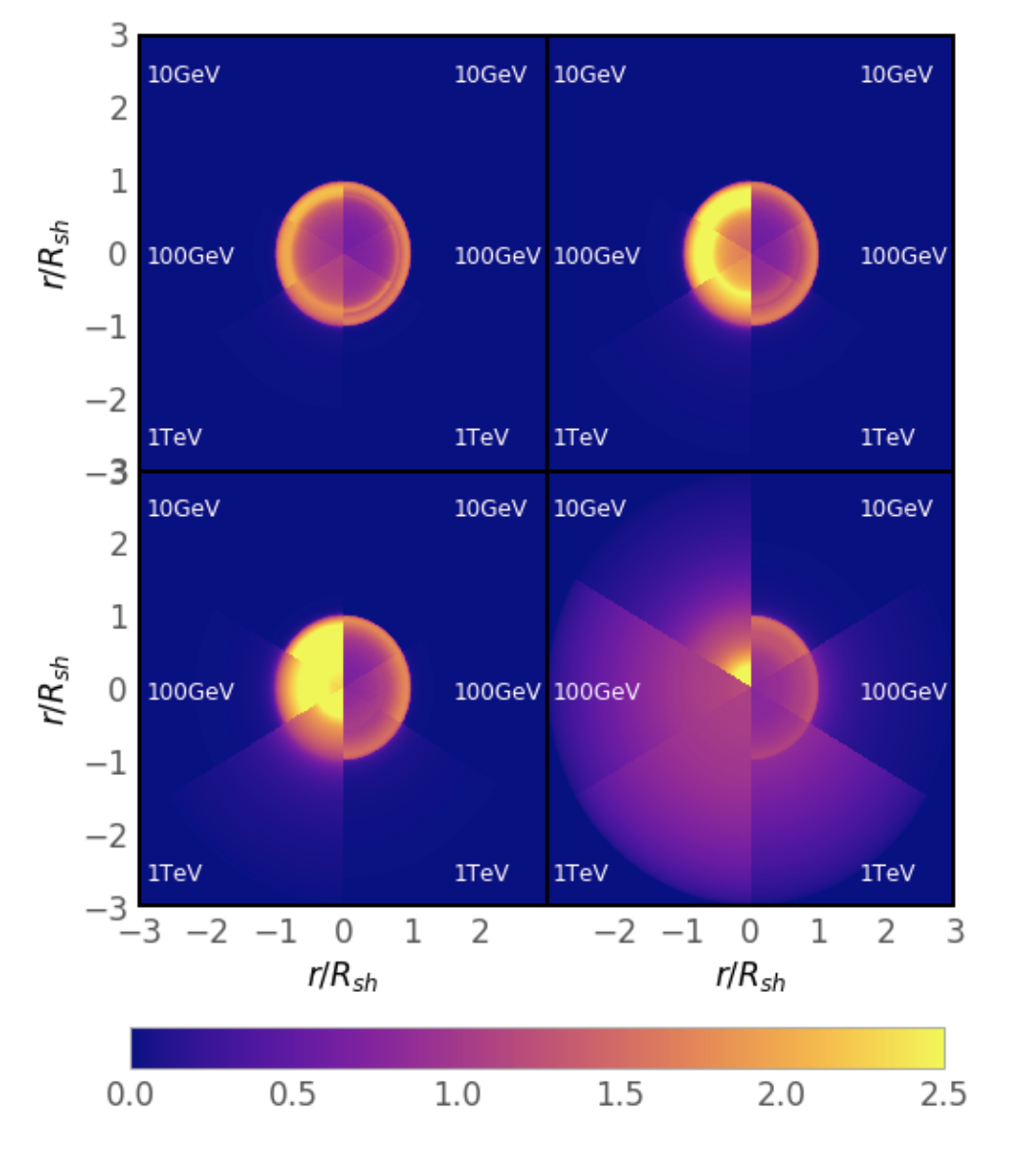}%grid
    \put(22,93){\color{white} IC}
    \put(32,93){\color{white} PD}
    \put(58,93){\color{white} IC}
    \put(68,93){\color{white} PD}
    \put(29.5,100){\color{white}\line(0,-1){78}}
    \put(65,100){\color{white}\line(0,-1){78}}
    \put(25,63){\color{white} 300\,yr}
    \put(62,63){\color{white} 1\,kyr}
    \put(25,27){\color{white} 5\,kyr}
    \put(62,27){\color{white} 10\,kyr}
    \end{overpic}
    \caption{Radial extension of the gamma-ray emission at four instants during SNR evolution. At each time, the left hemisphere of the SNR shows the inverse Compton (IC) emission, whilst the right hemisphere shows the pion decay (PD), with gamma-rays of different energies shown in sectors along the vertical axis. At 10\,kyr, an SNR halo has formed in both IC and PD scenarios, yet is more pronounced in the leptonic case. Morphological differences at earlier stages are also apparent \cite{BroseHalo}. }
    \label{fig:snr_halos}
\end{figure}

\subsection{Halos around Galactic Sources}
\label{sec:halos}

Energetic particles escape from their accelerator gradually, with the energy threshold for particle escape decreasing over time. In the immediate vicinity of the accelerator, yet situated in the interstellar medium (ISM) beyond the shock, there is an overabundance of energetic particles, forming a `halo' around the source. These halos may form around PWNe or SNRs (see sections \ref{sec:pwne} \& \ref{sec:snrs}) and typically have a CR density decreasing with increasing distance from the accelerator.

Figure \ref{fig:snr_halos} shows simulations for the formation of a halo around an SNR, with noticeable differences between the expected morphology and halo extent in the two cases of leptonic inverse Compton (IC) emission, and hadronic pion decay (PD) emission \cite{BroseHalo}. The halo effect is most pronounced at the oldest stage shown, of 10\,kyr; is more extended for higher energy particles (at these ages $\sim10$\,kyr); and more extended for IC than PD emission. However, it is an open question whether this difference is sufficient to distinguish IC from PD based on the halo morphology alone.

A similar phenomenon can be seen around PWNe, with several prominent halos recently detected around middle-aged pulsars. Although it is well-established that the diffusive transport of particles within PWNe is slow, on the order of $\sim 2\times10^{27}$cm$^2$s$^{-1}$ \cite{Tang,Porth}; in halos of escaped particles that have formed around older pulsars ($>100$\,kyr, rather than the $\sim1-10$\,kyr typical of PWNe), these escaped particles have effectively entered the ISM and can thus act as a probe of the surrounding ISM properties. 
Due to the low surface brightness, so far only halos that are comparatively close to Earth ($d\lesssim1$\,kpc) have been studied in depth \cite{HAWCGeminga,GiacintiHalo}. The canonical examples are the halos around the Geminga and Monogem pulsars, where the diffusion properties of the immediate ISM were inferred by HAWC to be a factor $\times 100$ lower than the expectation from the B/C ratio, at $4.5\times10^{27}$cm$^{2}$/s at 100\,TeV  \cite{HAWCGeminga}. 
%quote also the B/C value if a good reference can be found
%https://journals.aps.org/prl/pdf/10.1103/PhysRevLett.117.231102 (AMS)

This may be explained by some CR self-confinement effect, leading to suppression of the diffusion coefficient due to increased magnetic turbulence in the immediate vicinity of the accelerator, which several studies have shown for pulsar environments \cite{Evoli}, and \cite{BroseHalo} demonstrated that a similar effect is seen in SNRs. 
Alternative explanations have also been sought, such as turbulence of the surrounding medium generated by the shock wave of the parent SNR \cite{Fang}. A study presented in \cite{Recchia} investigated whether a ballistic (of the gyro-centre) component to the particle transport description may alleviate the requirement of a slow diffusion coefficient, finding that a more typical value of $D(1\mathrm{GeV})\sim10^{28}\mathrm{cm}^{-2}\mathrm{s}^{-1}$ can be kept,  although a highly efficient conversion of the pulsar spin-down energy into leptons is required in this case ($\sim60\%$ compared to $0.3-3\%$ for Geminga).
%\red{general expression for diffusion somewhere?}

Several experimental contributions presented measurements of the halos around the Geminga and Monogem pulsars, including detections by LHAASO and Tibet-AS$\gamma$ \cite{LHAASOGeminga,TibetGeminga}, an updated measurement by HAWC \cite{HAWCGeminga}, and the detection of extended emission around the Geminga pulsar by HESS, an IACT experiment \cite{HESSGeminga}. 
In the latter case, the gamma-ray emission from such a nearby source has an angular extent of $\sim5^\circ$ radius, well beyond the IACT field-of-view of $\lesssim 5^\circ$ in diameter, which makes such sources challenging for IACTs to analyse and detect.\footnote{The larger fields-of-view of e.g. $\sim10^\circ$ planned for Small-Sized Telescopes of the future CTA may alleviate this issue somewhat.}
Multiple contributions were presented on studies looking at improving the capabilities of current generation IACTs to highly extended gamma-ray sources \cite{MRMHona,VeritasChromey,HESSGeminga}. 

%No defined edge to halos

\subsection{Molecular Clouds as Cosmic Ray Tracers}
\label{sec:clouds}

As the distance $r$ between CRs and accelerators increases, the concentration of CRs (leptonic or hadronic) decreases as $r^{-\alpha}$ (where $\alpha$ depends on the particle acceleration and transport processes), such that the halo of escaping CRs is no longer detectable in gamma-rays beyond a certain distance. However, particularly in the case of hadronic CRs, they may remain sufficiently energetic to penetrate nearby interstellar clouds of material and interact to produce gamma-ray emission through pion decay once more. 
This effect was investigated in a modelling study \cite{CloudPeVModel} that identified plausible pairs of known SNRs and interstellar clouds, with the aim of predicting the gamma-ray flux from the clouds that should be detectable under the assumption that SNRs acted as PeVatrons in the past. As the most energetic particles escape at earlier times, they will not remain in the vicinity of the SNR for long, such that it is not surprising that we do not see evidence of active SNR PeVatrons. Signatures of past PeVatron activity may therefore be detectable from nearby clouds rather than from the parent SNR. A further aspect contributing to the low rate of observed SNR PeVatrons is that acceleration to PeV energies may only be achieved by a small $\sim1\%$ subset of the overall SN rate \cite{Cristofari}

CRs that have travelled sufficiently far from the original accelerator become isotropised in Galactic magnetic fields, forming the CR sea. This sea of CRs can also be probed by searching for evidence of enhanced gamma-ray emission from giant molecular clouds \cite{PeronCloud}, although in contrast to the aforementioned study, in this case the CRs are a `passive' sea, rather than an `active' flux of CRs arriving from a nearby accelerator. 
\cite{PeronCloud} showed evidence for enhanced GeV emission from GMCs in the 4-6\,kpc range, pointing towards an overabundance of CRs in this region, although the varying deviations from different GMCs discourage a purely radial dependence or global shift in the Galactic CR spectrum. 

\begin{figure}
    \centering
    \includegraphics[width=0.6\textwidth]{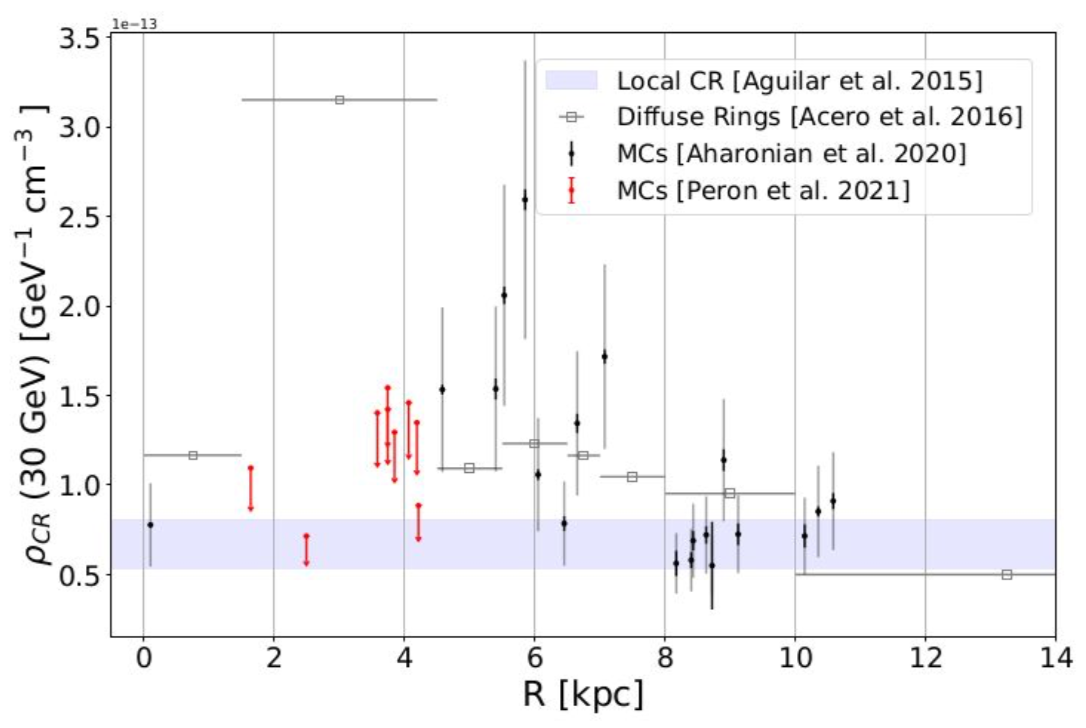}
    \caption{Measured GeV flux from selected Giant Molecular Clouds at different Galacto-centric distances, compared to the prediction for a homogeneous diffuse CR sea. An excess is seen most notably in the $\sim4-6$\,kpc region \cite{PeronCloud}. }
    \label{fig:GMC_gev}
\end{figure}

Conversely, an unusually low density of CRs was measured in the Central Molecular Zone (CMZ) \cite{CMZ}, indicating that some type of effective barrier against penetration of the CMZ by the CR sea may be at work in this region. 

\subsection{Stellar Clusters, Superbubbles and Winds}
\label{sec:windbubbles}

Given the theoretical and observational challenges with supernova remnants as the source class responsible for galactic CRs beyond the `knee', recently increased interest has turned towards stellar clusters and superbubbles as sources of CRs. 
Stellar clusters, such as Westerlund\,1, Westerlund\,2 and NGC\,3603, are large groups of stars of similar age; young stellar clusters in particular can harbour many recent supernovae and regions of active star formation.  %\red{check}. 

An analysis of Westerlund\,1 with the H.E.S.S. experiment showed variation in flux across the region - yet a spatially resolved spectral analysis found no statistically significant evidence for variation in spectral index \cite{West1}. Analyses of the Westerlund\,2 and NGC\,3603 regions with Fermi-LAT investigated the spatial coincidence of the gamma-ray emission to the stellar cluster and molecular material, finding a best-fit position coincident with the stellar cluster in both cases \cite{West2,NGCSaha}. For both Westerlund\,1 and Westerlund\,2  an hadronic model is preferred for the origin of the gamma-ray emission. 

Superbubbles are large, under-dense regions that have been evacuated due to the combined effect of multiple supernovae and/or stellar winds. 
The Cygnus superbubble situated in the Northern sky exhibits significant gamma-ray emission reaching above 100\,TeV; in fact, the highest energy photon recorded from the region by LHAASO had an energy of $1.42\pm0.13$\,PeV \cite{CygnusLHAASO}. Within the GeV range, studies using Fermi-LAT showed that multiple components with different morphologies, namely a Gaussian component and a template based on free-free emission in the region, likely contribute to the observed gamma-ray emission from the Cygnus region, being a good description of both the complex morphology and the spectral energy distribution \cite{CygnusXan}. %The total spectral energy distribution can be described by contributions from these different components to the complex morphology and total emission \cite{CygnusXan} \red{better description of study}. 
GeV data from Fermi-LAT and TeV data from HAWC were found to have a constant gamma-ray luminosity with radius, with the possible exception of the innermost GeV region, that can be well described by models of both a uniform ISM and a model including clouds randomly distributed along the line of sight \cite{CygnusMenchiari}. 
Further measurements of the Cygnus superbubble were reported by the HAWC and Tibet-AS$\gamma$ experiments, with gamma-ray emission above 100\,TeV detected in both cases. 
It remains unclear if the enhanced gamma-ray emission observed from stellar clusters and superbubbles is simply due to their constituents (SNRs, PWNe etc.) or if there are additional acceleration processes taking place, such as due to the combined stellar wind interactions. %\red{check and cite}

\subsection{Galactic Diffuse Emission}
\label{sec:diffuse}

%\red{Especially at high energies, explanation in terms of CR spectra}

\begin{figure}
    \centering
    \includegraphics[width=0.49\textwidth]{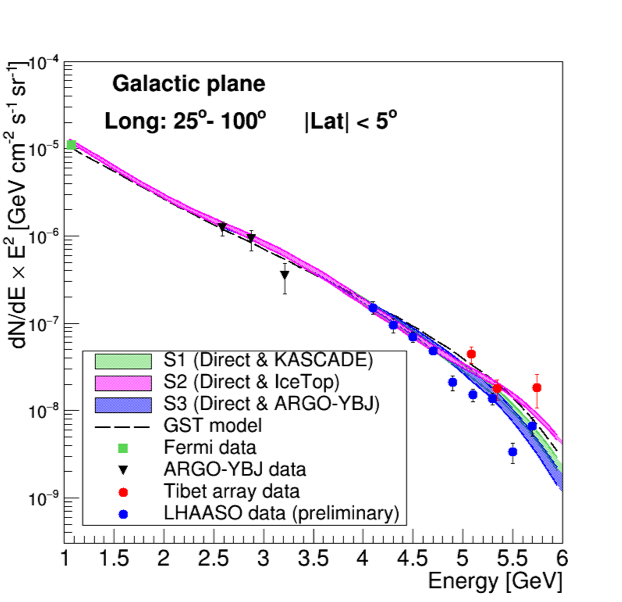}
    \includegraphics[width=0.49\textwidth]{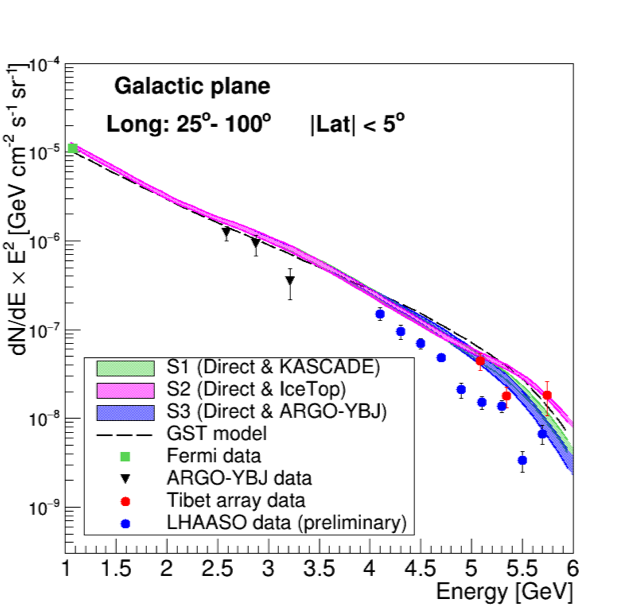}
    \caption{Comparison between measurements of the Galactic diffuse emission, including the preliminary LHAASO data first shown at this ICRC, and two different models for the CR distribution. Left: adopting the same spectrum for CRs throughout the Galaxy, Right: with a hardening of the CR spectrum towards the Galactic centre \cite{ModelGDE}. }
    \label{fig:GCR_model}
\end{figure}

Measurements of the galactic diffuse emission at TeV energies are important to determine the distribution of CRs within our Galaxy, as well as to understand its contribution as a background signal to weak sources or signatures of dark matter. %cite?
Both LHAASO and Tibet-AS$\gamma$ presented measurements of galactic diffuse emission in the sub-PeV band \cite{LHAASOGDE,TibetGDE}. These are shown on the same plot in figure \ref{fig:GCR_model}, together with two alternative models for the CR distribution in the Galaxy \cite{ModelGDE}. The first of the two models, mildly preferred by the LHAASO data, assumes the same spectrum for CRs throughout the Galaxy, whilst the second model, mildly preferred by the Tibet-AS$\gamma$ data, adopts a hardening of the CR spectrum towards the Galactic centre. Therefore, this indicates a possible tension between the two datasets, although it is not yet confirmed as significant given the uncertainties of the two measurements. 

\begin{figure}
    \centering
    \includegraphics[width=0.49\textwidth]{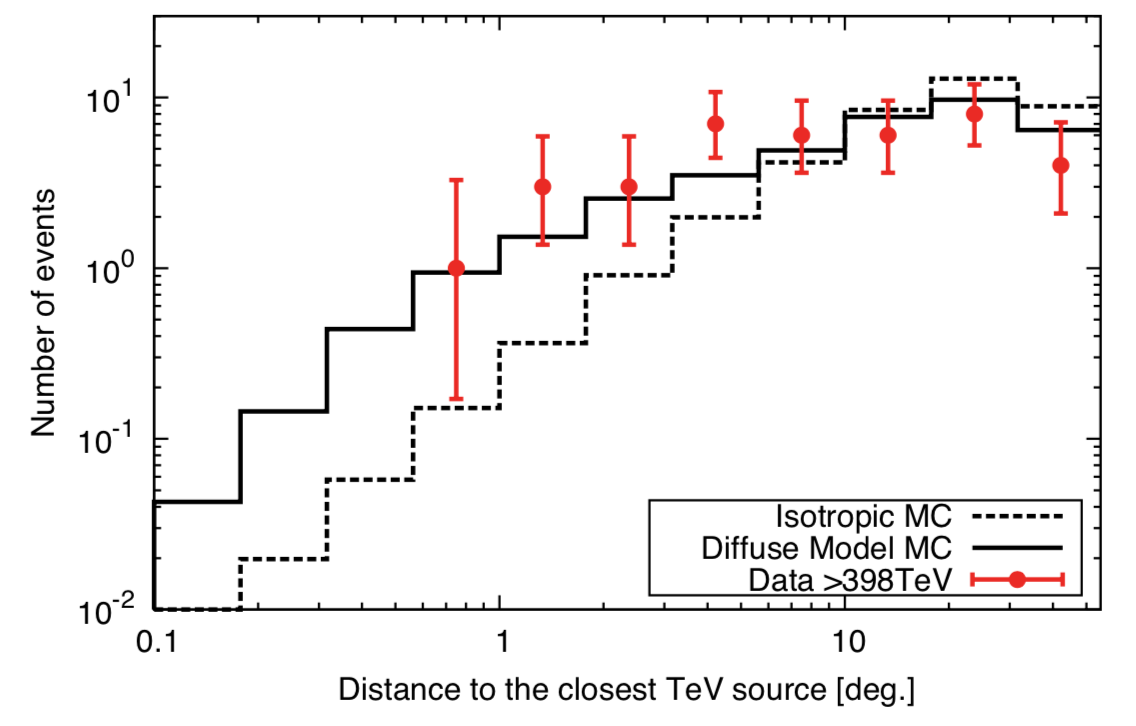}
    \includegraphics[width=0.40\textwidth]{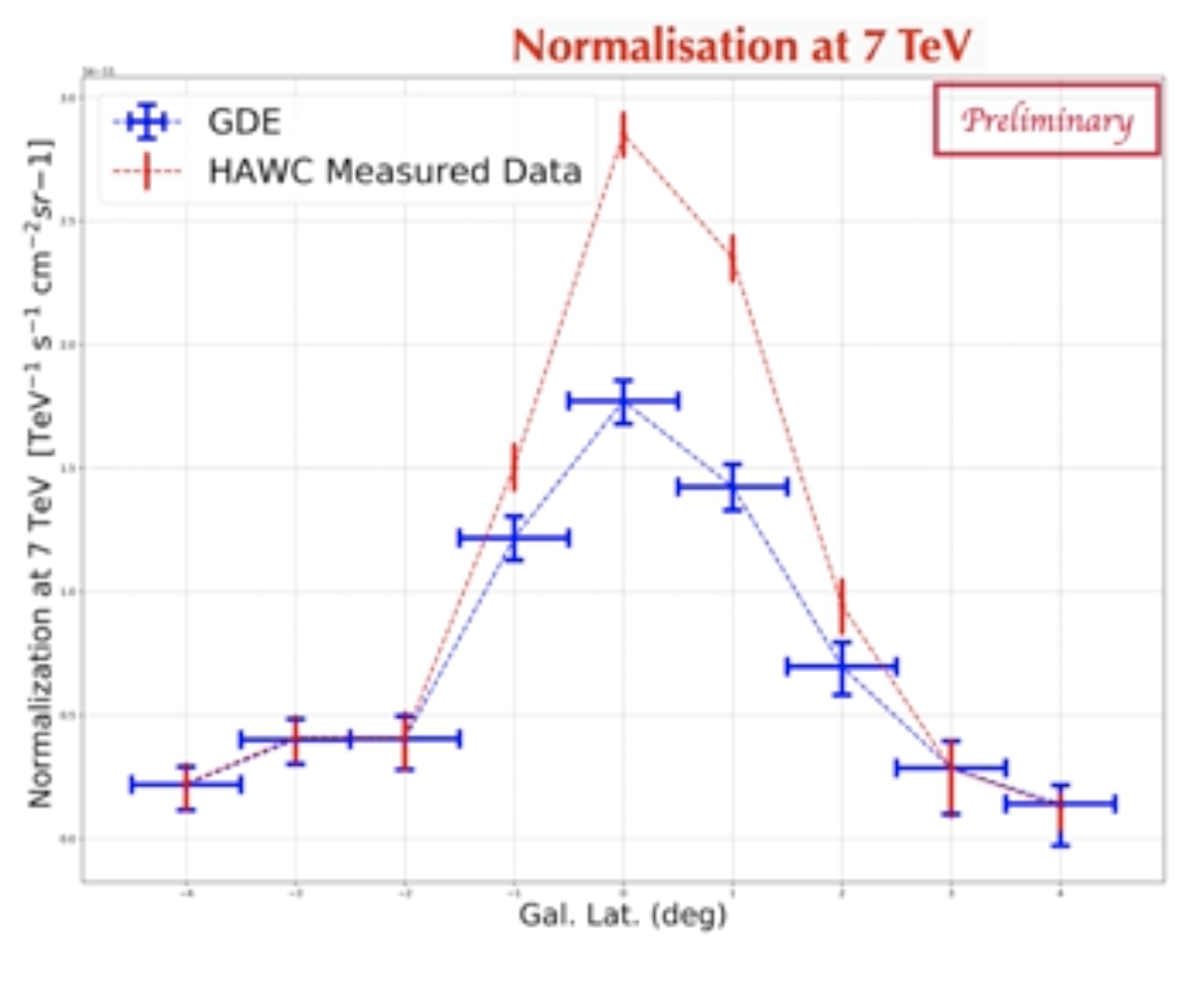}
    \caption{Measurements of the Galactic diffuse emission by (left) Tibet-AS$\gamma$ at energies $>398$\,TeV and (right) by HAWC, normalised at 7\,TeV \cite{TibetGDE,HAWCGDE}. }
    \label{fig:GCR_diffuse}
\end{figure}

The Tibet-AS$\gamma$ sub-PeV measurement is compared in the left panel of figure \ref{fig:GCR_diffuse} to models for Galactic diffuse emission and for an all-sky isotropic CR flux; the data is more consistent with the Galactic diffuse emission model. 
For any such diffuse emission measurement, known TeV sources must be excluded from the analysis - however, as yet unresolved sources may provide a non-negligible contribution to the measured diffuse emission. 
This may be the case for the HAWC diffuse emission measurement in the right panel of figure \ref{fig:GCR_diffuse}, where the data is seen to considerably exceed the model especially at low Galactic latitudes $|b|<2^\circ$ \cite{HAWCGDE}. Unresolved TeV sources in the Galactic plane may account for at least some of this discrepancy, although it is also worth noting that there are considerable uncertainties inherent in the models themselves. 

\section{Gamma-ray Bursts at Very-High-Energies}
\label{sec:GRBs}

Searches for the detection of gamma-ray bursts (GRBs) at VHE from ground-based facilities have been ongoing for several decades and are a prime target for current generation IACTs. 
It took until 2018 for the first detection, however, with a further three GRBs detected at VHE since. A summary of these GRBs is given in table \ref{tab:GRB}, all of which are long GRBs detected during the afterglow phase. A discussion session \textit{Gamma-Ray Bursts in the VHE regime} was held, dedicated to the observations and interpretation of these GRBs. 

The detection of VHE emission has opened several questions: i) Is VHE emission common in GRBs and what is its origin? ii) What is the maximum energy reached in GRBs? iii) Can the emission be described by a pure synchrotron component, or is an additional Synchrotron Self-Compton (SSC) component needed to explain the data? iv) Are there multiple regions of emission? 

Extragalactic gamma-ray sources, such as GRBs, are subject to redshift dependent absorption by the Extragalactic Background Light (EBL). Energetic gamma-ray photons undergo scattering interactions with ambient photons that leads to pair-production and absorption of the gamma-ray photon. This effect needs to be corrected for in order to obtain the intrinsic spectrum at the source location; yet models of the EBL are highly uncertain \cite{EBLreview}.  

\begin{table}
    \centering
    \begin{tabular}{lcclccc}
    GRB & Redshift ($z$) & $E_{\textrm{iso}}$ (erg) & Detected by & Obs. start time & Obs. duration & Ref.\\
    \hline 
    180720B & 0.654 & $6\times 10^{53}$ & H.E.S.S. & $T_0+10.1$h & 2h & \cite{GRB180720B} \\
    190114C & 0.4245 & $3\times 10^{53}$ & MAGIC & $T_0+57$s & 4.12h & \cite{GRB190114C} \\
    190829A & 0.08 & $2-5\times 10^{50}$ & H.E.S.S. & $T_0+4.3$h & 3.6h\,$\dagger$ & \cite{GRB190829A} \\
    201216C & 1.1 & $4.7\times 10^{53}$ & MAGIC & $T_0+56$s & 2.2h & \cite{GRB2012}
    \end{tabular}
    \caption{Summary of the four GRBs so far detected by IACTs. GRB identifiers are given by the date on which the GRB occurred, in YYMMDDX format. $T_0$ is the start time of the GRB event; $E_{\textrm{iso}}$ is the total energy output in gamma-rays assuming an isotropic energy release. \\
    $\dagger$ The information on H.E.S.S. observations provided in table \ref{tab:GRB} for GRB190829A refers to the first night only. Observations continued at $T_0+27.2$h for 4.7h and at $T_0+51.2$h for 4.7h, with significant emission from the direction of the GRB on all three nights. }
    \label{tab:GRB}
\end{table}

GRB 190114C was detected within $\sim100$\,s of the burst $T_0$ by MAGIC during observations taken under moonlight conditions \cite{GRB1901}.  
The spectrum of GRB 190114C provides indications for an additional SSC component being required to explain the VHE emission, although there are uncertainties inherent in GRB spectra, due to the EBL correction, that do not yet exclude alternative interpretations.

GRB 190829A was a very close GRB ($z=0.08$) observed over three consecutive nights by H.E.S.S., during which time the GRB energy flux was observed to decrease with the same slope in both the VHE and X-ray bands, implying a common origin for the two - such as a single synchrotron component \cite{HESSGRB}. GRB 201216C is the farthest GRB yet detected at VHE, with $z\sim1.1$, and was detected at more than 6\,sigma by MAGIC \cite{GRB2012}. 

\begin{figure}
    \centering %grid
    \begin{overpic}[width=0.6\textwidth]{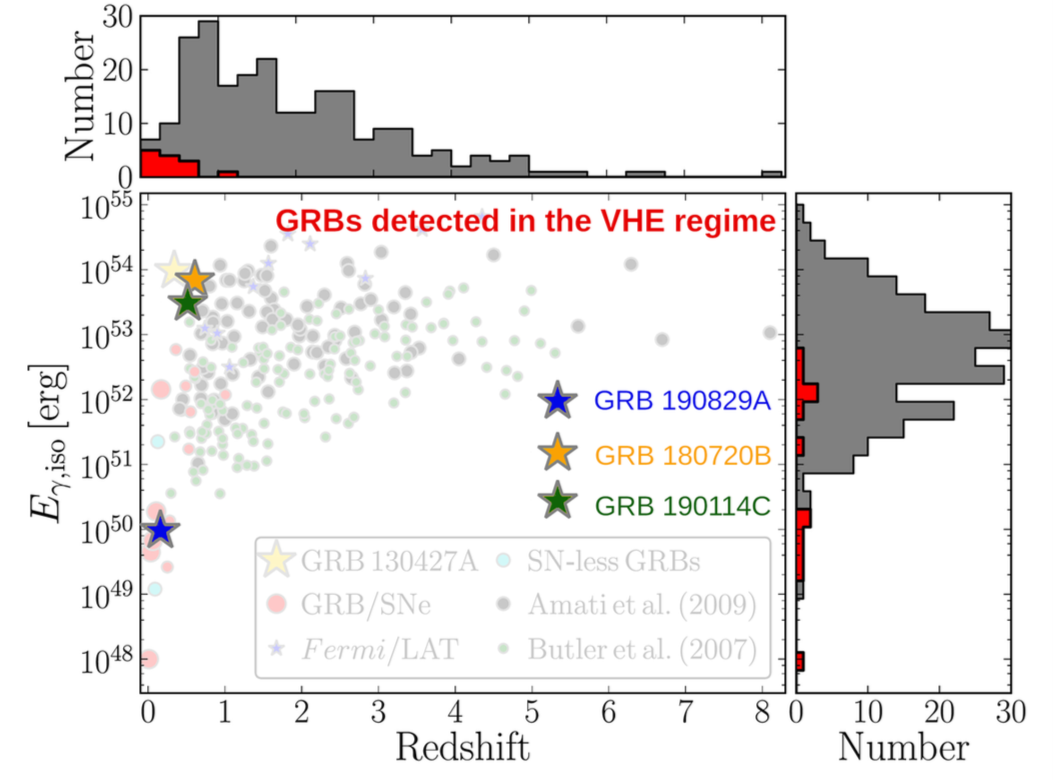}
    \put(20.5,44){$\bigstar$}
    \put(50.5,39){$\bigstar$ \scriptsize{GRB 201216C} }
    \end{overpic}
    \caption{Population of GRBs, with the four VHE GRBs indicated. Adapted from \cite{Levan}. }
    \label{fig:GRBpop}
\end{figure}

Figure \ref{fig:GRBpop} compares the VHE GRBs to the known GRB population. The four detected so far do not appear unusual either in redshift or in isotropic equivalent energy $E_{\gamma,\rm{iso}}$, implying that as typical examples, many more VHE GRB events can be anticipated in the future. 

A question naturally arose in discussion, that if these GRBs are so comparatively ordinary, why are the VHE detections only occurring recently? Several points could account for this. Based on lessons learnt in previous searches for GRBs, adjustments had been made to both the observation strategy and the detectors themselves. Hardware upgrades and improvements to alert pipelines were able to reduce the reaction time and improve the instrument sensitivity. Meanwhile, changes to the observation strategy also yielded dividends, by observing for a longer time after the initial burst (e.g. GRB 180720B at $>10$\,hours \cite{GRB180720B}) and conducting observations under moonlight conditions (e.g. GRB 190829A at $\sim 6\times$ the dark sky NSB \cite{GRB190114C}). 

The era of GRB observations in the VHE band is only just beginning; in particular, we can look forward to future instruments with continuous monitoring capabilities and an improved sensitivity thresholds increasing the VHE GRB population considerably.

\section{Summary and Outlook}
\label{sec:end}
%summary / science highlights

Given the considerable amount of contributions received to the two gamma-ray tracks at the 37$^{\rm th}$ ICRC, it is not possible to mention all in the limited time and space available to the rapporteurs. By reorganising the division between `direct' and `indirect' tracks along lines of scientific topic, we hope to have provided a more cohesive overview of the current status of the field, balancing breadth and depth. 

As far as Galactic gamma-ray astronomy is concerned, considerable focus is placed on the origin of Galactic Cosmic Rays, especially towards PeV energies corresponding to the knee of the CR spectrum. Recent results have brought the number of sources known to emit above 100\,TeV to 14 -- hence we find ourselves entering the PeV era. These so-called PeVatrons - accelerators of particles to PeV energies (using definition 2 of section \ref{sec:pevatron}) - will undoubtedly be subjected to close scrutiny in the coming years. As ground-based instruments such as HAWC, LHAASO and Tibet-AS$\gamma$ continue to gather data, we can look forward to updated measurements and further discoveries at the highest energies. 
A crucial next step to understand in the hunt for evidence of PeV CRs is the particle escape process - here considerable emphasis was placed both experimentally and theoretically on understanding halos of escaped particles (around both PWNe and SNRs), the diffuse sea of CRs, and on particle interactions with clouds. 

Lastly, an exciting area of development is that of transient events such as gamma-ray bursts. Indeed, the dynamic gamma-ray sky will continue to harbour surprises - a glimpse of which has already been provided in the elapsed time between the conclusion of this years ICRC in July and the conclusion of these proceedings in September. 

Until the next ICRC in two years time, we can look forward to progress in the next generation of IACT facilities, with the construction of CTA in particular imminently planned at both Northern and Southern sites. Currently operational ground-based gamma-ray facilities are located predominantly in the Northern hemisphere (see figure \ref{fig:GAI_worldmap}). 
A general sentiment shared by aficionados of ground-based gamma-ray facilities is a desire to observe the Southern sky and hence locate future particle detector experiments (such as SWGO) accordingly. 
With each technological improvement and new generation of experimental facilities, new discoveries are made - some questions will be answered, yet new questions will arise. Therefore we can look forward to exciting times ahead for VHE gamma-ray astronomy. 
%with a new experimental era just beginning. 

\section*{Acknowledgements}
%Include DFG here 
\noindent AM is supported by the Deutsche Forschungsgemeinschaft (DFG, German Research Foundation) – Project Number 452934793, MI 2787/1-1.

\end{document}